\documentclass[11pt]{article}
\pdfoutput=1
\usepackage{graphicx}
\usepackage{amssymb,amsmath}
\usepackage{cite,color,url}
\usepackage[colorlinks=true
,urlcolor=blue
,anchorcolor=blue
,citecolor=blue
,filecolor=blue
,linkcolor=blue
,menucolor=blue
,linktocpage=true
,pdfproducer=medialab
,pdfa=true
]{hyperref}

\usepackage{epsfig,psfrag,rotating,soul}
\usepackage{rotfloat}

\evensidemargin \oddsidemargin
\allowdisplaybreaks

\begin{document}
\thispagestyle{empty}

\def\thefootnote{\fnsymbol{footnote}}

\begin{flushright}
IFT-UAM/CSIC-13-024\\
\end{flushright}

\vspace{0.5cm}

\begin{center}

{\large\sc {\bf 
Non-decoupling SUSY in LFV Higgs decays:\\[.5em]
a window to new physics at the LHC}}

\vspace{1cm}

{\sc
M.~Arana-Catania$^{1}$%
\footnote{email: {\tt \href{mailto:miguel.arana@uam.es}{miguel.arana@uam.es}}}%
, E.~Arganda$^{2}$%
\footnote{email: {\tt \href{mailto:ernesto.arganda@fisica.unlp.edu.ar}{ernesto.arganda@fisica.unlp.edu.ar}}}%
, M.J.~Herrero$^{1}$%
\footnote{email: {\tt \href{mailto:maria.herrero@uam.es}{maria.herrero@uam.es}}}%
}

\vspace*{.7cm}

{\sl
$^1$Departamento de F\'isica Te\'orica and Instituto de F\'isica Te\'orica,
IFT-UAM/CSIC\\
Universidad Aut\'onoma de Madrid, Cantoblanco, Madrid, Spain

\vspace*{0.1cm}

$^2$IFLP, CONICET - Dpto. de F\'{\i}sica, Universidad Nacional de La Plata, \\ 
C.C. 67, 1900 La Plata, Argentina

}

\end{center}

\vspace*{0.1cm}

\begin{abstract}
\noindent
The recent discovery of a SM-like Higgs boson at the LHC, with a mass around 
125-126 GeV, together with the absence of results in the direct searches 
for supersymmetry, is pushing the SUSY scale ($m_\text{SUSY}$) into the 
multi-TeV range. This discouraging situation from a low-energy SUSY point 
of view has its counterpart in indirect SUSY observables which present 
a non-decoupling behavior with $m_\text{SUSY}$. This is the case of the 
one-loop lepton flavor violating Higgs decay rates induced by SUSY, which 
are shown here to remain constant as $m_\text{SUSY}$ grows, for 
large $m_\text{SUSY} >$ 2 TeV values and for all classes of intergenerational 
mixing in the slepton sector, $LL$, $LR$, $RL$ and $RR$. 
In this work we focus on the LFV decays of the three neutral MSSM Higgs 
bosons $h$, $H$, $A \to \tau \mu$, considering the four 
types of slepton mixing ($\delta_{23}^{LL}$, $\delta_{23}^{LR}$, 
$\delta_{23}^{RL}$, $\delta_{23}^{RR}$), and show that all 
the three channels could be measurable at the LHC, being $h \to \tau \mu$ 
the most promising one, with up to about hundred of events expected with 
the current LHC center-of-mass energy and luminosity. 
The most promising predictions for the future LHC stage are also included.

\end{abstract}

\def\thefootnote{\arabic{footnote}}
\setcounter{page}{0}
\setcounter{footnote}{0}

\newpage

\section{Introduction}
\label{intro}

The absence of any experimental signal, so far, in the searches for supersymmetry (SUSY) at the LHC~\cite{SUSYsearches} and the discovery of a new Higgs-like particle by ATLAS~\cite{Aad:2012tfa} and CMS~\cite{Chatrchyan:2012ufa} with a mass $m_{H_\text{SM}} \simeq$ 125-126 GeV, are pushing the SUSY mass parameters above the 1-TeV range. On one hand, the present lower mass bounds for squarks of the first and second generations and for gluinos are already above 1 TeV, and on the other hand, if the observed Higgs boson is identified with the lightest Higgs boson $h$ of the Minimal Supersymmetric Standard Model (MSSM), a radiatively corrected mass $m_{h} \simeq$ 125-126 GeV also implies rather heavy squark masses of the third generation (mainly stop masses) at or larger than 1 TeV. In principle, to place the SUSY masses at the multi-TeV range seems discouraging, both from an experimental point of view due to the inability to detect SUSY directly, and from a theoretical point of view, in regard to the naturalness of the theory, which contrarily suggests a soft SUSY-breaking scale, $m_{\rm SUSY}$, at or below the TeV scale. However, leaving the naturalness issue aside, the MSSM scenarios with very heavy SUSY masses can have other interesting aspects~\cite{heavySUSY}. In particular, this is the case of specific Higgs boson observables, like certain Higgs partial decay widths, which present a non-decoupling behavior with $m_{\rm SUSY}$, as shown, for instance, in~\cite{Haber:2000kq,Dobado:2001mq,Curiel:2002pf,Curiel:2003uk,Arganda:2004bz}, opening a new window to the indirect detection of heavy SUSY. As it is well known, the decoupling of SUSY radiative corrections in the asymptotic large SUSY mass limit is valid for SUSY theories with an exact gauge symmetry, in agreement with the general decoupling behavior of heavy states in Quantum Field Theory as stated in the famous Appelquist-Carazzone theorem~\cite{Appelquist:1974tg}. Nevertheless, it is also known that this theorem does not apply to theories with spontaneously broken gauge symmetries, nor with chiral fermions, which is the case of the MSSM. Furthermore, in order to have decoupling, the dimensionless couplings should not grow with the heavy masses. Otherwise, the mass suppression induced by the heavy-particle propagators can be compensated by the mass enhancement provided by the interaction vertices, with an overall non-decoupling effect, which is exactly what happens in some MSSM Higgs boson decays to fermions. For instance, it was studied in~\cite{Haber:2000kq} how non decoupling appears for large SUSY masses in the $h \to b \bar b$ decay through one-loop SUSY-QCD corrections, when the involved SUSY particle masses grow simultaneously with a generic soft SUSY-breaking scale $m_{\rm SUSY}$ (see also~\cite{Dobado:2001mq}). A similar non-decoupling behavior was obtained for flavor changing neutral Higgs boson decays into quarks of the second and third generations through both SUSY-QCD corrections~\cite{Curiel:2002pf} and SUSY-EW corrections~\cite{Curiel:2003uk}. Other interesting non-decoupling SUSY-EW effects have also been seen in lepton flavor violating (LFV) Higgs boson decays within the context of the MSSM-seesaw model~\cite{Arganda:2004bz}; and in Higgs-mediated LFV processes like: $\tau \to 3\mu$ decays~\cite{Arganda:2005ji}, some semileptonic $\tau$ decays~\cite{Arganda:2008jj,Herrero:2009tm} and in $\mu-e$ conversion in heavy nuclei~\cite{Arganda:2007jw}; all of them within the MSSM-seesaw model too. Other non-decoupling effects from heavy SUSY particles have also been noticed within the context of the SUSY inverse-seesaw model~\cite{InverseSeesaw}. Some studies of SUSY non decoupling within the MSSM have also been performed in the effective field theory approach where the effective Higgs-fermion-fermion vertices are induced from one-loop corrections of heavy SUSY particles~\cite{Dobado:2001mq,EffectiveVertices} and in $b \bar b h$ production~\cite{Liu:2012qu}. 

In the present work, motivated by the recent discovery of the SM-like Higgs boson, we will focus on the study of the LFV Higgs boson decays and we will work within the context of the MSSM at one-loop order with the hypothesis of general slepton flavor mixing. We will extend the study to the three neutral Higgs bosons of the MSSM $h$, $H$ and $A$, considering the new Higgs-like particle to be the lightest Higgs boson $h$. In particular, we will study the LFV Higgs decays $h \to \tau \mu$, $H \to \tau \mu$ and $A \to \tau \mu$. This kind of processes provide an unique window into new physics due to the strong suppression of flavor violation in the 
Standard Model (SM), where the flavor mixing is induced exclusively by 
the Yukawa couplings of the corresponding fermion sector. This is specially 
interesting in the lepton sector in which the flavor mixing will be hugely 
suppressed because of the smallness of the lepton Yukawa couplings. 
Therefore, the discovery of any process which violates the lepton flavor 
number would be a clear signal of physics beyond the SM. LFV is nowadays 
a very active field which is being studied in different models, through 
several channels (for a review, see for instance \cite{Raidal:2008jk}): 
radiative decays $\tau \to \mu \gamma$, $\tau \to e \gamma$ and 
$\mu \to e \gamma$; 3-body decays $\tau \to 3\mu$, $\tau \to 3e$ and 
$\mu \to 3e$; $\mu-e$ conversion in nuclei; semileptonic $\tau$ decays; 
among others. The specific case of LFV Higgs decays within supersymmetric 
models has also deserved special attention in 
the literature~\cite{Arganda:2004bz,LFVHDworks}. Also encouraging results for
the reach of LFV Higgs decays at the LHC have been recently obtained in \cite{Blankenburg:2012ex} within the context of a general effective Lagrangian approach.

Our purpose here is to take advantage of the mentioned non-decoupling behavior with $m_\text{SUSY}$ in the LFV Higgs decay widths into charged leptons of different generations, $\Gamma (\phi \to l_il_j)$, with $i \neq j$ and $\phi=h,H,A$, in order to look for sizeable branching ratios which can give rise to detectable signals at the LHC. Here and from now on, $l_il_j$ with $i \neq j$ in the final state of the LFV decays refers to either $l_i {\bar l}_j$ or 
${\bar l}_i l_j$. In general, the radiatively corrected LFV Higgs couplings to leptons are proportional to the heaviest lepton mass involved, being this the reason why we select the $\tau \mu$ channel as the most promising one. In addition, the $\mu e$ channel is extremely restricted by the associated radiative decay, $\mu \to e \gamma$~\cite{Adam:2013mnn}, leaving us almost no room to move in the allowed parameter space of slepton flavor mixing, and driving us to extremely low rates, not measurable in any LHC detector.
The $\tau e$ channel, on the other hand, gives us very similar results to the $\tau \mu$ channel, and from an experimental point of view, the LHC sensitivity to the former should be equivalent to the latter~\cite{LHCsensitivity}. Therefore the results obtained along this work for the LFV Higgs decays into $\tau \mu$ are straightforwardly translated into the $\tau e$ channels.
For the present study we will focus then on the $\phi \to \tau \mu$ decays within the MSSM at one loop with general slepton flavor mixing between the second and the third generations, without assuming any particular source of flavor mixing. Consequently, this general slepton mixing will be parametrized in a model-independent way by four dimensionless parameters, $\delta^{LL}_{23}$, $\delta^{LR}_{23}$, $\delta^{RL}_{23}$ and $\delta^{RR}_{23}$. It is important to remark that our calculations will be made in the mass eigenstate basis for all MSSM particles, including both charged sleptons and sneutrinos with the corresponding exact diagonalization of the full charged slepton and sneutrino mass matrices, and therefore we will not use here the Mass Insertion Approximation (MIA). The most important constraints to these four slepton mixings come from the $\tau \to \mu \gamma$ searches~\cite{Aubert:2009ag}. A recent study on updated constraints to all these general slepton flavor mixings $\delta^{AB}_{ij}$ in \cite{Arana-Catania:2013nha} indicates that, indeed, the present bounds to BR$(\tau \to \mu \gamma)$ lead to constraints on the 23 mixings which for 500 GeV $\leq \, m_{\rm SUSY}\,\leq$ 1500 GeV and 5 $\leq\,\tan \beta \,\leq$ 60 are at $|\delta^{LL}_{23}|_{\rm max},|\delta^{LR}_{23}|_{\rm max},|\delta^{RL}_{23}|_{\rm max} \sim 
{\cal O}(10^{-2}-10^{-1})$, and $|\delta^{RR}_{23}|_{\rm max} \sim 
{\cal O}(10^{-1}-1)$, and for heavier SUSY masses lead to weaker bounds. We will check that by raising the SUSY mass scale into the multi-TeV range will turn into the needed relaxation of these bounds, since these LFV radiative decays present a decoupling behavior with the SUSY scale. This will allow us to explore a high SUSY mass scale region, where we will find very promising values for the LFV Higgs rates.

Our paper is structured as follows: in Section~\ref{theory} we describe the MSSM frame which we will use in the rest of the paper, introducing the flavor mixing and the scenarios we will work with. Section~\ref{LFVrates} is devoted to the LFV Higgs decay rates in the heavy SUSY particle mass limit, showing the non-decoupling behavior of these rates with $m_\text{SUSY}$. We dedicate Section~\ref{LFVxsections} to the relevant numerical results of the LFV Higgs event rates at the LHC, considering the effects induced by each type of delta, $\delta^{AB}_{23}$. This will be done taking into account the conditions of the present and future phases of the LHC, in order to estimate the number of events one could expect for each LFV Higgs channel. We will finally close the paper with the conclusions, contained in Section~\ref{conclusions}.

\section{The MSSM with general slepton mixing}
\label{theory}

In this paper we explore SUSY scenarios with general flavor mixing in the slepton sector, which have the same particle content as the MSSM. Within these scenarios, the most general hypothesis of flavor mixing assumes a non-diagonal mass matrix in flavor space, for both charged slepton and sneutrino sectors. The charged slepton mass matrix is a $6 \times 6$ matrix due to the six electroweak interaction eigenstates, ${\tilde l}_{L,R}$ with $l=e, \mu, \tau$. The inclusion, within the MSSM, of only three sneutrino eigenstates, ${\tilde \nu}_{L}$ with $\nu=\nu_e, \nu_\mu, \nu_ \tau$, reduces the sneutrino mass matrix to a $3 \times 3$ matrix. In the case in which slepton and sneutrino mass matrices were diagonal, we would still have a tiny flavor violation induced by the PMNS matrix of the neutrino sector and transmitted by the tiny neutrino Yukawa couplings, but we neglect it here.

The non-diagonal entries in the $6 \times 6$ slepton matrix can be described in a model-independent way in terms of a set of dimensionless parameters $\delta_{ij}^{AB}$ ($A,B=L,R$; $i,j=1,2,3$), where $L,R$ stand for the possible chiralities of the lepton partners and $i,j$ indexes run over the three generations. These scenarios with general sfermion flavor mixing lead generally to larger LFV rates than in the so-called Minimal Flavor Violation (MFV) scenarios, where the mixing is induced exclusively by the Yukawa coupling of the corresponding fermion sector. This statement is true for both squarks and sleptons but it is obviously of special interest in the slepton case due to the extremely small size of the lepton Yukawa couplings. Hence, in the present case of slepton mixing, the $\delta_{ij}^{AB}$'s will provide the unique origin of LFV processes with potentially measurable rates.

One usually starts with the non-diagonal $6 \times 6$ slepton squared mass matrix referred to the electroweak interaction basis, which we order here as ($\tilde e_{L}$, $\tilde \mu_{L}$, $\tilde \tau_{L}$, $\tilde e_{R}$, $\tilde \mu_{R}$, $\tilde \tau_{R}$), and write this matrix in terms of left- and right-handed blocks $M^2_{\tilde l \, AB}$ ($A,B=L,R$), which are non-diagonal $3\times 3$ matrices,
\begin{equation}
{\mathcal M}_{\tilde l}^2 =\left( \begin{array}{cc}
M^2_{\tilde l \, LL} & M^2_{\tilde l \, LR} \\ 
M_{\tilde l \, LR}^{2 \, \dagger} & M^2_{\tilde l \,RR}
\end{array} \right) \,,
\label{eq:slep-6x6}
\end{equation} 
 where
 \begin{eqnarray}\label{eq:slep-matrix}
 M_{\tilde l \, LL \, ij}^2 
&=& m_{\tilde L \, ij}^2 + \left( m_{l_i}^2
   + (-\frac{1}{2}+ \sin^2 \theta_W ) M_Z^2 \cos 2\beta \right) \delta_{ij} \,, \notag\\
 M^2_{\tilde l \, RR \, ij}
&=& m_{\tilde E \, ij}^2 + \left( m_{l_i}^2
   -\sin^2 \theta_W M_Z^2 \cos 2\beta \right) \delta_{ij} \notag \,, \\
 M^2_{\tilde l \, LR \, ij}
&=& v_1 {\cal A}_{ij}^l- m_{l_{i}} \mu \tan\beta \, \delta_{ij} \,,
 \end{eqnarray}
with flavor indexes $i,j=1,2,3$, corresponding to the first, second and third generations, respectively. $\theta_W$ is the weak angle, $M_Z$ is the $Z$ gauge boson mass, $(m_{l_1},m_{l_2}, m_{l_3})=(m_e,m_\mu,m_\tau)$ are the lepton masses, $\tan\beta=v_2/v_1$ with $v_1=\left< {\cal H}_1^0 \right>$ and $v_2=\left< {\cal H}_2^0 \right>$, the two vacuum expectation values of the corresponding neutral Higgs boson in the Higgs $SU(2)$ doublets, ${\cal H}_1= ({\cal H}^0_1\,\,\, {\cal H}^-_1)$ and ${\cal H}_2= ({\cal H}^+_2 \,\,\,{\cal H}^0_2)$, and $\mu$ is the usual higgsino mass term. It should be noted that the non diagonality in flavor comes exclusively from the soft SUSY-breaking parameters, which could be non vanishing for $i \neq j$, namely: the masses $m_{\tilde L \, ij}$ for the slepton $SU(2)$ doublets $(\tilde \nu_{Li}\,\,\, \tilde l_{Li})$, the masses $m_{\tilde E \, ij}$ for the slepton $SU(2)$ singlets $(\tilde l_{Ri})$, and the trilinear couplings, ${\cal A}_{ij}^l$.

In the sneutrino sector there is, correspondingly, a one-block $3\times 3$ mass matrix, respect to the $(\tilde \nu_{eL}, \tilde \nu_{\mu L}, \tilde \nu_{\tau L})$ electroweak interaction basis,
\begin{equation}
{\mathcal M}_{\tilde \nu}^2 =\left( \begin{array}{c}
M^2_{\tilde \nu \, LL}  
\end{array} \right),
\label{eq:sneu-3x3}
\end{equation} 
where
\begin{equation} 
 M_{\tilde \nu \, LL \, ij}^2 
 =  m_{\tilde L \, ij}^2 + \left( 
   \frac{1}{2} M_Z^2 \cos 2\beta \right) \delta_{ij} \,.
\label{eq:sneu-matrix}
\end{equation} 
 
It should be also noted that, due to $SU(2)_L$ gauge invariance, the same soft masses $m_{\tilde L \, ij}$ enter in both the slepton and sneutrino $LL$ mass matrices. Besides, in the previous equations we have neglected the neutrino mass terms, which due to their extremely small value are totally irrelevant for the present computation. 

The general slepton flavor mixing is then introduced via the non-diagonal terms in the soft-breaking slepton mass matrices and trilinear coupling matrices, which are defined here as
\noindent \begin{equation}
m^2_{\tilde L}= \left(\begin{array}{ccc}
 m^2_{\tilde L_{1}} & \delta_{12}^{LL} m_{\tilde L_{1}}m_{\tilde L_{2}} & \delta_{13}^{LL} m_{\tilde L_{1}}m_{\tilde L_{3}} \\
 \delta_{21}^{LL} m_{\tilde L_{2}}m_{\tilde L_{1}} & m^2_{\tilde L_{2}} & \delta_{23}^{LL} m_{\tilde L_{2}}m_{\tilde L_{3}}\\
\delta_{31}^{LL} m_{\tilde L_{3}}m_{\tilde L_{1}} & \delta_{32}^{LL} m_{\tilde L_{3}}m_{\tilde L_{2}}& m^2_{\tilde L_{3}} \end{array}\right) \,, \label{mLL}
\end{equation}
 \noindent \begin{equation}
v_1 {\cal A}^l =\left(\begin{array}{ccc}
m_e A_e & \delta_{12}^{LR} m_{\tilde L_{1}}m_{\tilde E_{2}} & \delta_{13}^{LR} m_{\tilde L_{1}}m_{\tilde E_{3}}\\
\delta_{21}^{LR} m_{\tilde L_{2}}m_{\tilde E_{1}} & m_\mu A_\mu & \delta_{23}^{LR} m_{\tilde L_{2}}m_{\tilde E_{3}}\\
\delta_{31}^{LR} m_{\tilde L_{3}}m_{\tilde E_{1}} & \delta_{32}^{LR} m_{\tilde L_{3}} m_{\tilde E_{2}}& m_{\tau}A_{\tau}\end{array}\right) \,, \label{v1Al}
\end{equation}
\noindent \begin{equation}
m^2_{\tilde E}= \left(\begin{array}{ccc}
 m^2_{\tilde E_{1}} & \delta_{12}^{RR} m_{\tilde E_{1}}m_{\tilde E_{2}} & \delta_{13}^{RR} m_{\tilde E_{1}}m_{\tilde E_{3}}\\
 \delta_{21}^{RR} m_{\tilde E_{2}}m_{\tilde E_{1}} & m^2_{\tilde E_{2}} & \delta_{23}^{RR} m_{\tilde E_{2}}m_{\tilde E_{3}}\\
\delta_{31}^{RR} m_{\tilde E_{3}} m_{\tilde E_{1}}& \delta_{32}^{RR} m_{\tilde E_{3}}m_{\tilde E_{2}}& m^2_{\tilde E_{3}} \end{array}\right) \,. \label{mRR}
\end{equation}

In all this work, for simplicity, we are assuming that all $\delta_{ij}^{AB}$ parameters are real. Therefore, hermiticity of ${\mathcal M}_{\tilde l}^2$ and
${\mathcal M}_{\tilde \nu}^2$ implies $\delta_{ij}^{AB}= \delta_{ji}^{BA}$.
Besides, in order to avoid extremely large off-diagonal matrix entries we restrict ourselves to $|\delta_{ij}^{AB}| \leq 1$. It is worth to have in mind for the rest of this work that our parametrization of the off-diagonal in flavor space entries in the above mass matrices is purely phenomenological and does not rely on any specific assumption about the origin of the MSSM soft-mass parameters. In particular, it should be noted that our parametrization for the $LR$ and $RL$ squared mass entries connecting different generations (i.e. for $i \neq j$) assumes a similar generic form to the $LL$ and $RR$ entries. For instance, $M^2_{\tilde l \, LR \, 23}= \delta_{23}^{LR} m_{\tilde L_{2}}m_{\tilde E_{3}}$. 
This implies that our hypothesis for the trilinear off-diagonal couplings ${\cal A}^l_{ij}$ with $i \neq j$ (as derived from Eq.(\ref{v1Al})) is one among other possible definitions considered in the literature. In particular, it is related to the usual assumption $M^2_{\tilde l \, LR \, ij} \sim v_1 m_{\rm SUSY}$, by setting ${\cal A}^l_{ij} \sim {\cal O}(m_{\rm SUSY})$. Here, $v_1^2+v_2^2=v^2=2\frac{M_W^2}{g^2}$, $M_W$ is the charged gauge boson mass, $g$ is the $SU(2)_L$ gauge coupling  and $m_{\rm SUSY}$ is a typical SUSY mass scale.  It should be also noted that the diagonal entries in Eq.(\ref{v1Al}) have been normalized as it is usual in the literature, namely, by factorizing out the corresponding lepton Yukawa coupling: ${\cal A}^l_{ii}= y_{l_i} A^l_{ii}$, with $A^l_{11}=A_e$, $A^l_{22}=A_\mu$ and $A^l_{33}=A_\tau$. Finally, it should be mentioned that our choice in Eqs.(\ref{mLL}), (\ref{v1Al}) and (\ref{mRR}) is to normalize the non-diagonal in flavor entries with respect to the geometric mean of the corresponding diagonal squared soft masses. For instance, 
\begin{eqnarray}
&&\delta^{LL}_{23}= (m^2_{\tilde L})_{23}/(m_{\tilde L_{2}}m_{\tilde L_{3}}), \,\,\,\,\, 
\delta^{RR}_{23}= (m^2_{\tilde R})_{23}/(m_{\tilde R_{2}}m_{\tilde R_{3}}),
~~~\nonumber\\
&&\delta^{LR}_{23}= (v_1 {\cal A}^l)_{23}/(m_{\tilde L_{2}}m_{\tilde R_{3}}), \,\,\,\,\,
\delta^{RL}_{23}=\delta^{LR}_{32}= (v_1 {\cal A}^l)_{32}/(m_{\tilde L_{3}}m_{\tilde R_{2}}).
\label{deltas23defs}
\end{eqnarray}

The next step is to rotate the sleptons and sneutrinos from the electroweak interaction basis to the physical mass eigenstate basis, 
\begin{eqnarray}
\left( \begin{array}{c} \tilde l_{1} \\ \tilde l_{2} \\ \tilde l_{3} \\
                                \tilde l_{4}   \\ \tilde l_{5}  \\\tilde l_{6}   \end{array} \right)
   =  R^{\tilde l} \left( \begin{array}{c} \tilde e_L \\ \tilde \mu_L \\ \tilde \tau_L \\ 
  \tilde e_R \\ \tilde \mu_R \\ \tilde \tau_R \end{array} \right) ~,~~~~
\left( \begin{array}{c}  \tilde \nu_{1} \\ \tilde \nu_{2}  \\  \tilde \nu_{3}  \end{array} \right)              =  R^{\tilde \nu}  \left( \begin{array}{c} \tilde \nu_{eL} \\ \tilde \nu_{\mu L}  \\  \tilde \nu_{\tau L}   \end{array} \right) ~,
\label{newsquarks}
\end{eqnarray} 
with $R^{\tilde l}$ and $R^{\tilde \nu}$ being the respective $6\times 6$ and $3\times 3$ unitary rotating matrices which yield the diagonal mass-squared matrices as follows,
\begin{eqnarray}
{\rm diag}\{m_{\tilde l_1}^2, m_{\tilde l_2}^2, 
          m_{\tilde l_3}^2, m_{\tilde l_4}^2, m_{\tilde l_5}^2, m_{\tilde l_6}^2 
           \}  & = &
R^{\tilde l}   {\cal M}_{\tilde l}^2    
 R^{\tilde l \dagger}    ~, \label{sleptons}\\
{\rm diag}\{m_{\tilde \nu_1}^2, m_{\tilde \nu_2}^2, 
          m_{\tilde \nu_3}^2  
          \}  & = &
R^{\tilde \nu}     {\cal M}_{\tilde \nu}^2    
 R^{\tilde \nu \dagger}    ~.
\label{sneutrinos} 
\end{eqnarray}

In the numerical computations of the present study we will restrict ourselves to the case where there are flavor mixings exclusively between the second and third generations of sleptons, thus we set all $\delta_{ij}^{AB}$'s to zero except for $ij=23$. On one hand, the LFV one-loop corrected Higgs couplings are proportional to the heaviest lepton mass involved~\cite{Arganda:2004bz} and, therefore, the Higgs decay rates into $\mu e$ are suppressed by a factor $m_\mu^2/m_\tau^2$ with respect to the $h, H, A \to \tau \mu, \tau e$ decay rates. On the other hand, the related LFV radiative decay $\mu \to e \gamma$ has a much more restrictive upper bound~\cite{Adam:2013mnn} than $\tau \to e \gamma$ and $\tau \to \mu \gamma$ decays~\cite{Aubert:2009ag}, and the present allowed values of the $\delta_{12}^{AB}$'s would not drive us to any measurable $\phi \to \mu e$ rates. 

In order to simplify our analysis, and to reduce further the number of independent parameters, we will focus on the following numerical study on simplified SUSY scenarios, where the relevant soft-mass parameters are related to a single SUSY mass scale, $m_\text{SUSY}$. In particular we choose the following setting for the relevant mass parameters:
\begin{eqnarray}
m_{\tilde L} &=& m_{\tilde E} = m_\text{SUSY} \,, \\ \label{msleptons}
\mu &=& M_2 = a \,m_\text{SUSY} \,, \label{M2mu}
\end{eqnarray}
where $a$ is a constant coefficient that we will fix in the next sections to different values, namely, $a = \frac{1}{5}$, $\frac{1}{3}$, 1.
We also set an approximate GUT relation for the gaugino masses:
\begin{equation}
M_2 = 2 M_1 = M_3/4 \,.
\end{equation}
Here and in the following we use a short notation for the common soft masses, namely, $m_{\tilde L}$ for $m_{\tilde L}=m_{\tilde L_{1}}=m_{\tilde L_{2}}=m_{\tilde L_{3}}$, etc. For simplicity, we have also assumed vanishing soft-trilinear couplings for the first and second generations in the slepton sector, i.e., $A_\mu=A_e=0$. We have checked that other choices with non-vanishing values for any of these two couplings do not alter the conclusions of this work. The trilinear coupling for the third generation has been fixed here to the generic SUSY mass scale, $A_\tau = m_\text{SUSY}$. 

Regarding the non-diagonal trilinear couplings we have also assumed a rather simple but realistic setting by relating them with the single soft SUSY-breaking mass scale, $m_\text{SUSY}$. Specifically, we assume the following linear relation:
\begin{eqnarray}
 {\cal A}^l_{23} &=& {\tilde \delta}^{LR}_{23} m_\text{SUSY} \,, \,\,\,\,\,  
 {\cal A}^l_{32} = {\tilde \delta}^{LR}_{32} m_\text{SUSY} \,,
 \label{trilinear23}
\end{eqnarray} 
where the new  dimensionless parameters ${\tilde \delta}^{LR}_{23}$ and ${\tilde \delta}^{LR}_{32}$ are trivially related to the previously introduced ones ${\delta}^{LR}_{23}$ and ${\delta}^{LR}_{32}$ of Eq.(\ref{deltas23defs}) by:
\begin{equation}
{\delta}^{LR}_{23}= \left( \frac{v_1}{m_\text{SUSY}} \right) {\tilde \delta}^{LR}_{23}\,, \,\,\,\,\,
{\delta}^{LR}_{32}= \left( \frac{v_1}{m_\text{SUSY}} \right) {\tilde \delta}^{LR}_{32}.
\label{LRversusLRtilde}
\end{equation}
It is clear from Eq.(\ref{LRversusLRtilde}) that ${\delta}^{LR}_{23}$ and ${\delta}^{LR}_{32}$  scale with $m_\text{SUSY}$ as $\sim \frac{1}{m_\text{SUSY}}$. Therefore, in the forthcoming analysis of the LFV observables, whenever  the decoupling behavior of these observables with large $m_\text{SUSY}$ be explored we will use instead the more suited parameters ${\tilde \delta}^{LR}_{23}$ and ${\tilde \delta}^{LR}_{32}$, which can be kept fixed to a constant value while taking large $m_\text{SUSY}$ values. 
 
Concerning the size of the flavor violating trilinear couplings that are of relevance here, ${\cal A}^l_{23}$ and ${\cal A}^l_{32}$, there are well-known theoretical upper bounds that arise from  vacuum stability. If any of these trilinear couplings is too large, the MSSM scalar potential develops a charge and/or color breaking (CCB) minimum deeper than the Standard-Model-like (SML) local minimum or an unbounded from below (UFB) direction in the field space \cite{Casas:1996de}. Then the requirement of the absence of 
these dangerous CCB minima or UFB directions implies specific upper bounds on the size of the non-diagonal trilinear couplings, and consequently on the size of the flavor changing deltas. For the case of interest here the upper bound from stability can be written simply as \cite{Casas:1996de}:
\begin{equation}
|{\cal A}^l_{23}| \leq y_{\tau} \sqrt{m_{\tilde L_{2}}^2+m_{\tilde E_{3}}^2+m_1^2},
\end{equation}  
and similarly for ${\cal A}^l_{32}$. 
Here, 
\begin{equation} 
y_{\tau}=\frac{gm_\tau}{\sqrt{2}M_W \cos \beta}
\end{equation} 
is the Yukawa coupling of the tau lepton, and the squared soft mass $m_1^2$ can be written as:
\begin{equation} 
m_1^2= (m^2_A+M_W^2+M_Z^2 \sin^2\theta_W) \sin^2\beta- \frac{1}{2}M_Z^2.
\end{equation}  
In our simplified scenarios for the slepton, gaugino and Higgs sectors, with just three MSSM input parameters,  $m_\text{SUSY}$, $\tan\beta$, and $m_A$, and by considering Eqs.(\ref{trilinear23}) and (\ref{LRversusLRtilde}), the previous bound implies in turn the following bound on ${\delta}^{LR}_{23}$, and correspondingly on 
${\tilde \delta}^{LR}_{23}$ (and similar bounds for $2 \leftrightarrow 3$):
\begin{equation}  
| {\delta}^{LR}_{23}|\leq \frac{m_\tau}{m_\text{SUSY}} \sqrt{2+\frac{m_1^2}{m^2_\text{SUSY}}}\,\,,\,\,\,
|{\tilde \delta}^{LR}_{23}| \leq y_{\tau}\sqrt{2+\frac{m_1^2}{m^2_\text{SUSY}}}.
\label{delta23bounds}
\end{equation} 
For example, if we take $m_\text{SUSY}=m_A=1$ TeV we get upper bounds for $|{\tilde \delta}^{LR}_{23}|$ of ${\cal O}(0.1)$ in the low $\tan\beta$ region close to 5, and of ${\cal O}(1)$ in the large $\tan\beta$ region close to 50. These correspond to an upper bound on  $|{\delta}^{LR}_{23}|$ of $\sim 0.0035$ that is nearly independent on $\tan\beta$  and it gets weaker for larger $m_\text{SUSY}$ values due to the scaling factor 
($\frac{m_\tau}{m_\text{SUSY}}$) in Eq.(\ref{delta23bounds}).

However, the reliability of these bounds have been questioned in the literature because of the fact that the existence of deeper minima than the SML local minimum does not necessarily imply a problem whenever the lifetime of this false SML vacuum is sufficiently long. In this later case, other theoretical upper bounds based on metastability then apply. Indeed, by demanding that the lifetime of the whole observable universe staying at the SML vacuum be longer than the age of the universe the constraints on the flavor changing deltas get much more relaxed~\cite{Park:2010wf}. According to  \cite{Park:2010wf} the upper bounds on ${\delta}^{LR}_{23}$ from metastability, in contrast to the limits from stability, turn out to be independent on the Yukawa coupling, they do not decouple for asymptotically large ${m_\text{SUSY}}$ and they are dependent on $\tan\beta$. For instance, for $\tan\beta= 3$ and ${m_\text{SUSY}}=5$ TeV the metastability limit on ${\delta}^{LR}_{23}$ gets weaker than its stability bound by a factor of 40, whereas for $\tan\beta= 30$ and ${m_\text{SUSY}}=5$ TeV it is weaker by a smaller factor of 4, leading to approximate upper bounds of $|{\delta}^{LR}_{23}|\leq 0.02$ and $|{\delta}^{LR}_{23}|\leq 0.002$ respectively. This translates into an upper bound of  
about $|{\tilde \delta}^{LR}_{23}| \leq 2-3$ for $3 \leq \tan\beta \leq 30$ and 
${m_\text{SUSY} }\leq 10$ TeV.
The numerical estimates of these metastability bounds are done considering each delta separately (i.e setting the other deltas to zero value) and for a specific assumption of the relevant Euclidean action providing the decay probability via tunneling of the metastable vacuum into the global minimum. Changing the input value for the Euclidean action can increase notably the maximum allowed value up to almost doubling it, leading to roughly $|{\tilde \delta}^{LR}_{23}| \leq 4-6$. The effects on these bounds from switching on more than one delta at the same time have not been considered yet in the literature but they could relevantly modify these bounds. For the present work, and given the uncertainty in all these estimates of the upper bounds from stability and metastability arguments, we will consider a rather generous interval when performing the numerical estimates of branching ratios and event rates. Concretely we will choose several examples for ${\tilde \delta}^{LR}_{23}$ of very different size that will be taken within the wide interval $|{\tilde \delta}^{LR}_{23}| \leq 10$. This corresponds to $|{\delta}^{LR}_{23}|\leq 0.009$ for the particular values of $m_\text{SUSY} =$ 5 TeV  and $\tan\beta =$ 40.

On the other hand, the values of the soft masses of the squark sector are irrelevant for LFV processes, except in the present case of LFV MSSM Higgs bosons decays where these parameters enter in the prediction of the radiatively corrected Higgs boson masses. Since we want to identify the discovered boson with the lightest MSSM Higgs boson, we will set these parameters to values which give a prediction of $m_h$ compatible with the LHC data. Specifically, we fix them to the particular values $m_{\tilde Q} =$ $m_{\tilde U} =$ $m_{\tilde D} =$ $A_t =$ $A_b =$ 5 TeV, which we have checked provide a value for $m_h$ that lies within the LHC-favored range [121 GeV, 127 GeV] for all the MSSM parameter space considered here. 

In summary, the input parameters of our simplified SUSY scenarios are: the mass of the pseudoscalar Higgs boson, $m_A$, the ratio of the two Higgs vacuum expectation values, $\tan\beta$, the generic SUSY mass scale, $m_\text{SUSY}$, and the four delta parameters, $\delta_{23}^{LL}$, $\delta_{23}^{RR}$, $\delta_{23}^{LR}$ and $\delta_{23}^{RL}$ (or ${\tilde \delta}^{LR}_{23}$ and ${\tilde \delta}^{RL}_{23}$ alternatively to the two latter), which we vary within the following intervals:
\begin{itemize}
\item 200 GeV $\leq m_A \leq$ 1000 GeV,
\item 1 $\leq \tan\beta \leq$ 60,
\item 0.5 TeV $\leq m_\text{SUSY} \leq$ 10 TeV,
\item $-1 \leq$ $\delta_{23}^{LL}$, $\delta_{23}^{RR}$ $\leq 1$,
\item $-10 \leq$ ${\tilde \delta}_{23}^{LR}$, ${\tilde \delta}_{23}^{RL}$ $\leq 10$,\\
(or, equivalently, $|{\delta}^{LR}_{23}|, |{\delta}^{RL}_{23}|\leq 0.009$ for $m_\text{SUSY} =$ 5 TeV  and $\tan\beta =$ 40). 

\end{itemize}

\section{Results for the branching ratios of the LFV decays}
\label{LFVrates}
In this section we study the behavior of the radiative corrections from SUSY loops to the LFV neutral Higgs bosons decays in the heavy SUSY particle mass limit and compare them with the case of LFV radiative lepton decays. For the forthcoming estimates of the LFV Higgs decay rates, we use the complete one-loop formulas and the full set of diagrams contributing to the $\Gamma (\phi \to {\bar l_i} l_j)$ and $\Gamma (\phi \to l_i {\bar l_j})$ partial decay widths, with $i \neq j$, within the MSSM, which are written in terms of the mass eigenvalues for all the involved MSSM sparticles, including the physical slepton and sneutrino masses, $m_{\tilde l_i} \,\,(i=1,..,6)$ and $m_{\tilde \nu_i}\,\,(i=1,2,3)$, and the rotation matrices $R^{\tilde l}$ and $R^{\tilde \nu}$ of Eqs.~(\ref{sleptons}) and~(\ref{sneutrinos}). We take these general one-loop formulas from \cite{Arganda:2004bz} and emphasize that they are valid for the general slepton mixing case considered here, with all the mixing effects from the $\delta^{AB}_{ij}$'s being transmitted to the LFV Higgs decay rates via the physical slepton and sneutrino masses and their corresponding rotations. The off-diagonal trilinear couplings in Eq.(\ref{v1Al}) also enter into this computation of the LFV Higgs decay rates. For a more detailed discussion about these analytical results we refer the reader to~\cite{Arganda:2004bz}. It should be also noted that, since we are assuming real $\delta^{AB}_{ij}$'s, the predictions for BR$(\phi \to l_i {\bar l_j})$ and the CP-conjugate BR$(\phi \to {\bar l_i} l_j)$ are the same. Thus, we will perform our estimates for just one of them and will denote this rate generically by BR$(\phi \to l_il_j)$. Obviously, in the case that these two channels
cannot be differentiated experimentally one should then add the two contributions to the total final number of events. However, for the present computation we do not sum them and report instead the results for $\phi \to l_il_j$, meaning that they are valid for any of the two cases. 

As said above, we focus only on $h \to \tau \mu$, $H \to \tau \mu$ and $A \to \tau \mu$ decay channels and consider the constraints imposed over the parameter space by the current upper bound on the related LFV radiative decay $\tau \to \mu \gamma$~\cite{Aubert:2009ag}. The SUSY mass spectra are computed numerically with the code {\tt SPheno}~\cite{SPheno}. The slepton and sneutrino spectra are computed from an additional subroutine that we have implemented into {\tt SPheno} in order to include our parametrization of slepton mixing given by the 
$\delta^{AB}_{ij}$'s. 
The LFV decay rates are computed with our private FORTRAN code in which we have implemented the complete one-loop formulas for the LFV partial Higgs decay widths of~\cite{Arganda:2004bz} and the complete one-loop formulas for the LFV radiative $\tau$ decay widths which we take from~\cite{Arganda:2005ji}. Note that these latter formulas for the $\tau \to \mu \gamma$ decays are also written in terms of the physical sparticle eigenvalues and eigenstates and do not neglect any of the lepton masses. The mass spectrum of the MSSM Higgs sector, with two-loop corrections included, and their corresponding total widths are calculated by means of the code {\tt FeynHiggs}~\cite{FeynHiggs}. 

\begin{figure}[t!]
\begin{center}
\begin{tabular}{cc}
\includegraphics[width=80mm]{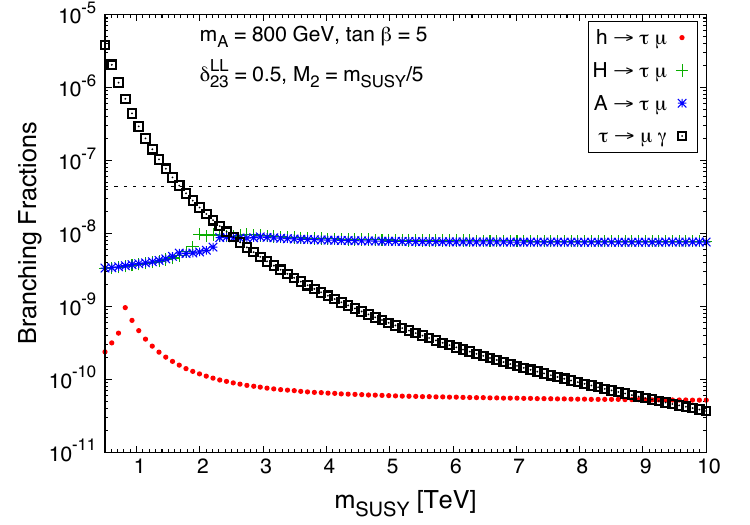} &
\includegraphics[width=80mm]{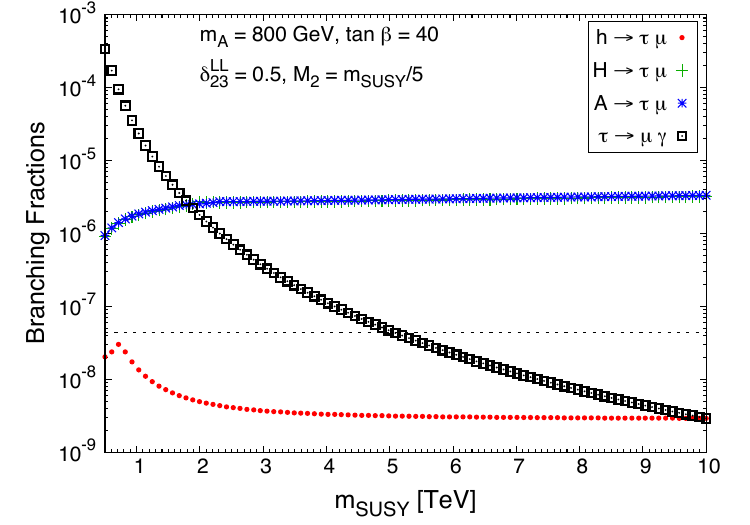} \\
\includegraphics[width=80mm]{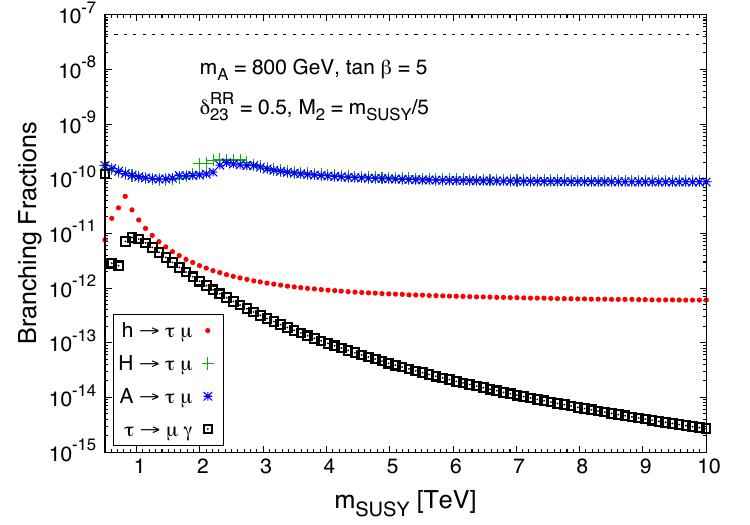} &
\includegraphics[width=80mm]{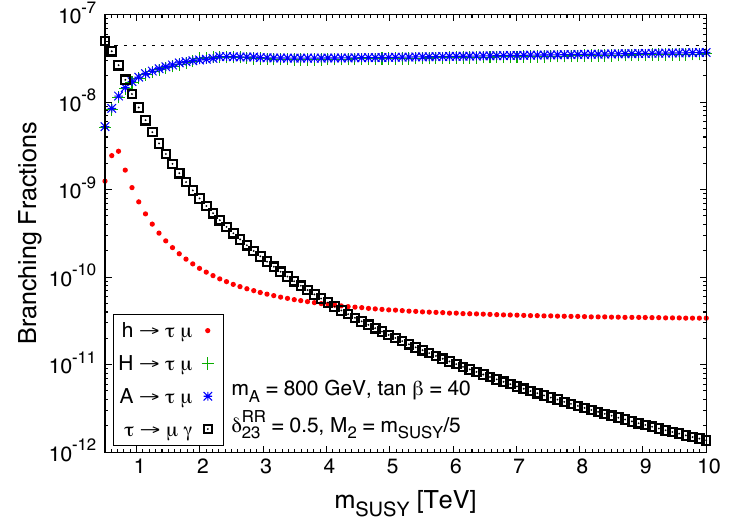} \\
\includegraphics[width=80mm]{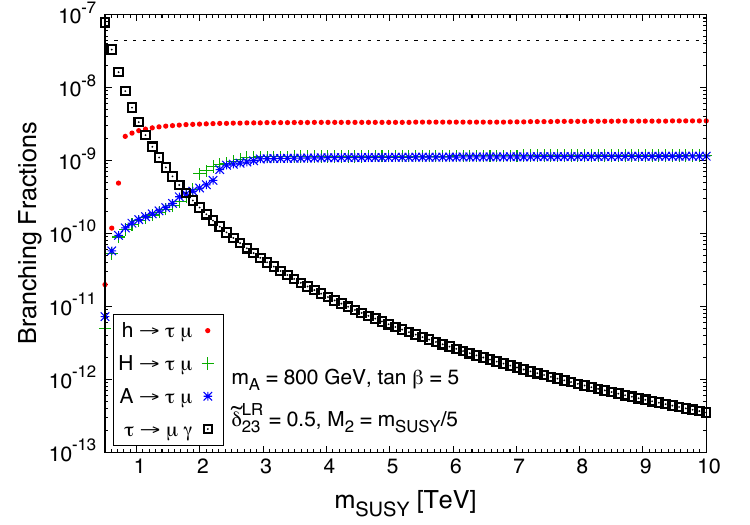} &
\includegraphics[width=80mm]{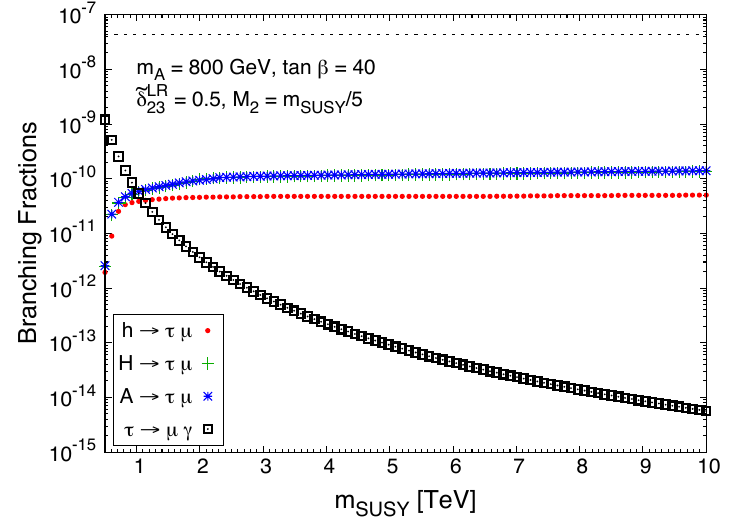}

\end{tabular}
\caption{Large $m_\text{SUSY}$ behaviour of the LFV decay rates: BR($h \to \tau \mu$), BR($H \to \tau \mu$), BR($A \to \tau \mu$) and 
BR($\tau \to \mu \gamma$) as functions of $m_\text{SUSY}$ for $\tan\beta = 5$ (left panels) and $\tan\beta = 40$ (right panels) with $\delta_{23}^{LL} = 0.5$ (upper panels), $\delta_{23}^{RR} = 0.5$ (middle panels) and $\tilde \delta_{23}^{LR} = 0.5$ (lower panels). The results for $\tilde \delta_{23}^{RL} = 0.5$ (not shown) are identical to those of $\tilde \delta_{23}^{LR} = 0.5$. In each case, the other flavor changing deltas are set to zero. In all panels, $m_A =$ 800 GeV and the other MSSM parameters are set to the values reported in the text, with $M_2 = \frac{1}{5} m_\text{SUSY}$. 
The horizontal dashed line denotes the current experimental upper bound for $\tau \to \mu \gamma$ channel, BR($\tau \to \mu \gamma$) $< 4.4 \times 10^{-8}$~\cite{Aubert:2009ag}.}\label{BRs-mSUSY}
\end{center}
\end{figure}

Next we present the numerical results for the branching ratios of the LFV decays. We show in Figure~\ref{BRs-mSUSY} the behavior of the branching ratios for the two types of LFV decays, BR$(\phi \to \tau \mu)$ and BR$(\tau \to \mu \gamma$), as functions of $m_\text{SUSY}$, and we consider two different values of $\tan\beta$, namely, $\tan\beta = 5$ (left panels) and $\tan\beta = 40$ (right panels). In each case we set one single delta to be non vanishing with the particular values: $\delta_{23}^{LL} = 0.5$ (upper panels), $\delta_{23}^{RR} = 0.5$ (middle panels) and ${\tilde \delta}_{23}^{LR} = 0.5$ (lower panels). All the other flavor changing deltas are set to zero. We find identical results 
for ${\tilde \delta}_{23}^{RL} = 0.5$ as for ${\tilde \delta}_{23}^{LR} = 0.5$ and, for brevity, we have omitted the plots for ${\tilde \delta}_{23}^{RL}$ in Figure~\ref{BRs-mSUSY}.   

\begin{figure}[t!]
\begin{center}
\begin{tabular}{cc}
\includegraphics[width=80mm]{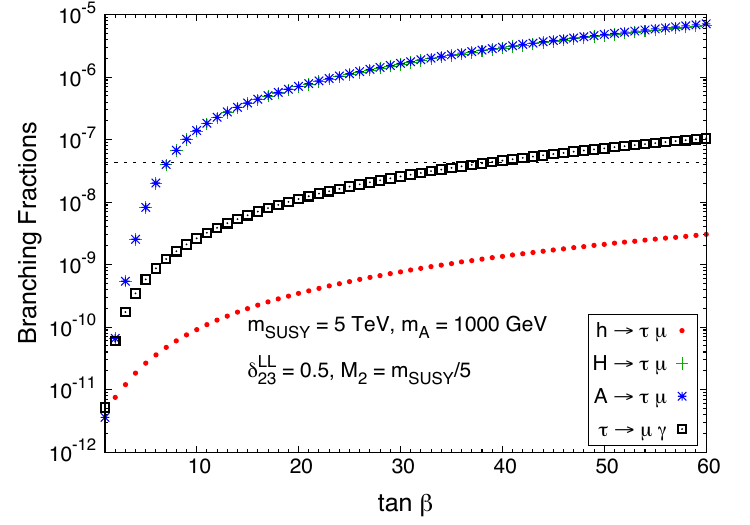} &
\includegraphics[width=80mm]{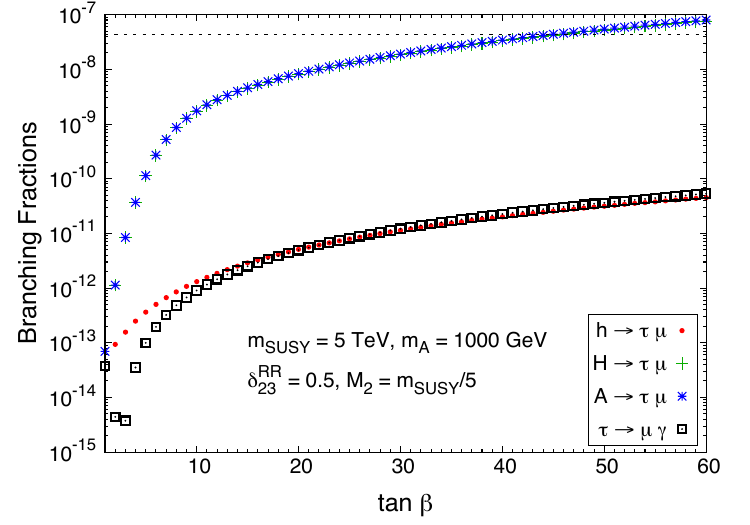} \\
\includegraphics[width=80mm]{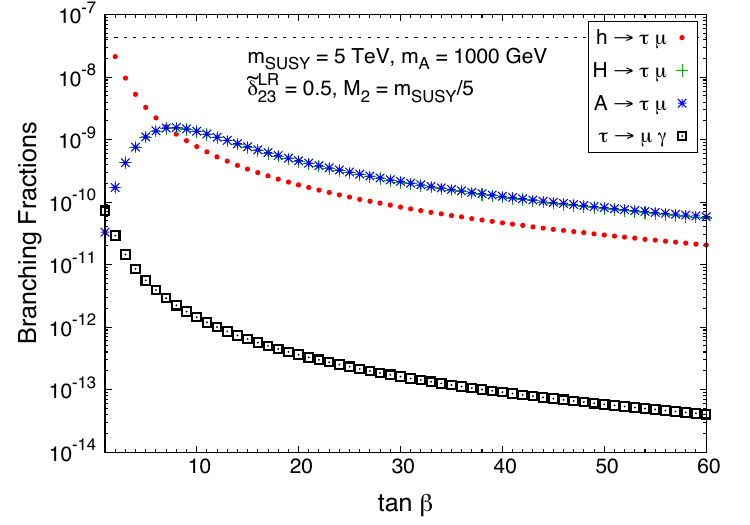} &
\includegraphics[width=80mm]{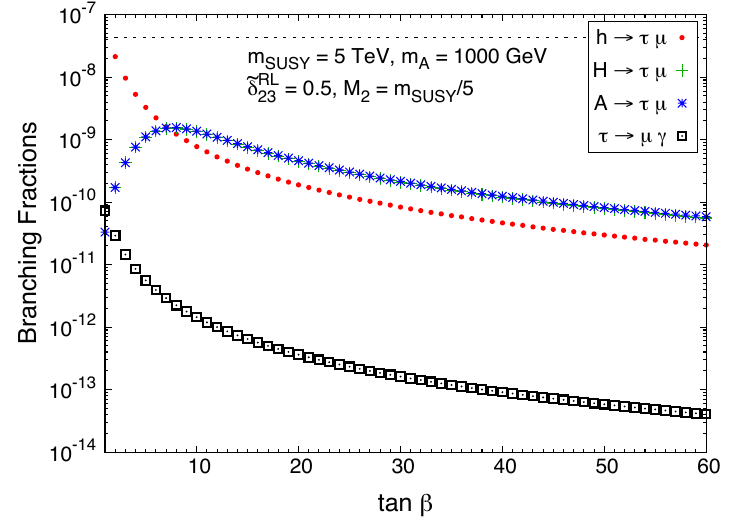}

\end{tabular}
\caption{BR($h \to \tau \mu$), BR($H \to \tau \mu$), BR($A \to \tau \mu$) and 
BR($\tau \to \mu \gamma$) as functions of $\tan\beta$ for $\delta_{23}^{LL} =$ 0.5 (upper left panel), $\delta_{23}^{RR} =$ 0.5 (upper right panel), $\tilde \delta_{23}^{LR} =$ 0.5 (lower left panel) and $\tilde \delta_{23}^{RL} =$ 0.5 (lower right panel). In each case, the other flavor changing deltas are set to zero. In all panels, $m_A =$ 1000 GeV, $m_\text{SUSY} =$ 5 TeV and the other MSSM parameters are set to the values reported in the text, with $M_2 = m_\text{SUSY}/5$. The horizontal dashed line denotes the current experimental upper bound for $\tau \to \mu \gamma$ channel, BR($\tau \to \mu \gamma$) $< 4.4 \times 10^{-8}$~\cite{Aubert:2009ag}.}\label{BRs-tanb}
\end{center}
\end{figure}

On the upper panels of Figure~\ref{BRs-mSUSY}, when the responsible for the flavor mixing between the second and the third generations is $\delta_{23}^{LL}$, the branching ratios of the LFV Higgs decays show a clear non-decoupling behavior with $m_\text{SUSY}$, which remain constant from $m_\text{SUSY} \simeq$ 1 TeV to $m_\text{SUSY} =$ 10 TeV, with values of BR$(h \to \tau \mu) \simeq 5 \times 10^{-11}$ and BR$(H, A \to \tau \mu) \simeq 8 \times 10^{-9}$ for $\tan\beta =$ 5, and BR$(h \to \tau \mu) \simeq 3 \times 10^{-9}$ and BR$(H, A \to \tau \mu) \simeq 3 \times 10^{-6}$ for $\tan\beta =$ 40. Another important feature that should be noted is the fast growing with $\tan\beta$ of these decays which increase the $H$ and $A$ LFV decay rates almost three orders of magnitude from $\tan\beta =$ 5 to $\tan\beta =$ 40. Furthermore, we have numerically checked that for large values of $\tan\beta \geq 10$ the partial decay widths $\Gamma(H, A \to \tau \mu)$ go approximately as $\sim (\tan\beta)^4$~\cite{Arganda:2004bz}, whereas $\Gamma(h \to \tau \mu)$ goes as $\sim (\tan\beta)^2$. This implies that the corresponding branching ratios go all at large $\tan\beta \geq 10$ as BR($h, H, A \to \tau \mu$) $\sim (\tan\beta)^2$ in this $LL$ case, since the total widths go as $\Gamma_\text{tot}(H, A) \sim (\tan\beta)^2$ and $\Gamma_\text{tot}(h)$ is approximately constant with $\tan\beta$. This behavior of the BRs with $\tan\beta$ is confirmed numerically in our forthcoming Figure~\ref{BRs-tanb}. In contrast, the branching ratio of the $\tau \to \mu \gamma$ decay presents a decoupling behavior with $m_\text{SUSY}$, decreasing as $\sim 1/m_\text{SUSY}^4$, and it is reduced around five orders of magnitude from $m_\text{SUSY} =$ 500 GeV to $m_\text{SUSY} =$ 10 TeV. In all these figures we have also included, for comparison, the experimental upper  bound for the $\tau \to \mu \gamma$ channel, whose present value is BR($\tau \to \mu \gamma$) $< 4.4 \times 10^{-8}$~\cite{Aubert:2009ag}. Thus, each $m_\text{SUSY}$ point which leads to a prediction of BR$(\tau \to \mu \gamma)$ above this line is excluded by data. Therefore, only values of $m_\text{SUSY} \gtrsim$ 2 TeV for $\tan\beta =$ 5 and $m_\text{SUSY} \gtrsim$ 5 TeV for $\tan\beta =$ 40, and their corresponding predictions for the LFV rates, are allowed for $\delta_{23}^{LL} = 0.5$.

On the middle panels of Figure~\ref{BRs-mSUSY} we have plotted the LFV Higgs and radiative decay rates as functions of $m_\text{SUSY}$, considering $\delta_{23}^{RR}$ as the responsible for $\tau-\mu$ mixing. A similar non-decoupling behavior to the $\delta_{23}^{LL}$ case can be observed for $\delta_{23}^{RR}$, whose branching ratios stay again constant as $m_\text{SUSY}$ grows. However, the numerical contribution of $\delta_{23}^{RR}$ to the LFV processes is much less important than that of $\delta_{23}^{LL}$, and all the $RR$ rates are in comparison around two orders of magnitude smaller than the $LL$ rates. In the $RR$ case, another important feature is that all the predictions found of BR($\tau \to \mu \gamma$) for $5 \leq \tan \beta \leq 40$ and $m_\text{SUSY}$ values above 500 GeV are allowed by the present BR$(\tau \to \mu \gamma)$ experimental upper bound. 

The predictions of BR($h \to \tau \mu$), BR($H \to \tau \mu$), BR($A \to \tau
\mu$) and BR($\tau \to \mu \gamma$) as functions of $m_\text{SUSY}$, for the
case ${\tilde \delta}_{23}^{LR} =$ 0.5, are shown on the lower panels of
Figure~\ref{BRs-mSUSY}. We see clearly again a non-decoupling behavior with
$m_\text{SUSY}$, since the branching ratios of the Higgs decays tend to a constant
value as $m_\text{SUSY}$ grows, in contrast to the BR$(\tau \to \mu \gamma)$
rates that display again a decoupling behavior and decrease with
$m_\text{SUSY}$. For this particular choice of  ${\tilde \delta}_{23}^{LR} =$
0.5 we also see that the predicted branching ratios for the LFV $h$ decays at low $\tan
\beta= 5$ are  larger than in the previous $LL$ and $RR$ cases, whereas the branching ratios
for the LFV $H$ and $A$ decays are larger than those of the $RR$ case but
smaller than the $LL$ ones.  The lower right panel shows that for
$\tan\beta=40$ the branching ratios of the three Higgs bosons, $h$, $H$ and $A$, are comparatively smaller than for $\tan\beta=5$. This decreasing with $\tan\beta$ of the LFV decay rates for the $LR$ case with fixed value of ${\tilde \delta}_{23}^{LR}$ is confirmed in our forthcoming  Figure~\ref{BRs-tanb}. In consequence, the largest LFV Higgs decay rates that will be obtained in the $LR$ (and $RL$) case will be for low $\tan \beta$ values and this will be taken into account in our next studies in order to maximize the event rates from these decays at the LHC.  

To sum up the main results of Figure~\ref{BRs-mSUSY}, the most relevant 
$\delta_{23}^{AB}$ parameter at low $\tan \beta$ values for the lightest Higgs boson $h$ is ${\tilde \delta}_{23}^{LR}$ (and ${\tilde\delta}_{23}^{RL}$), 
which gives rise to larger LFV Higgs decay rates than $\delta_{23}^{LL}$ 
and $\delta_{23}^{RR}$, whereas for the $H$ and $A$ Higgs bosons the most relevant parameter is 
$\delta_{23}^{LL}$. At large $\tan \beta$ values, the most relevant parameter for all the three Higgs bosons is  $\delta_{23}^{LL}$. All these branching ratios, as we will see later, can be further enhanced by
considering two non-vanishing deltas at the same time, by exploring with larger size of these deltas, and also by considering different signs for the various deltas. Overall, the main conclusion from this Figure~\ref{BRs-mSUSY} is that  
if one wants to obtain sizeable and 
allowed by data branching ratios, one needs large values of $m_\text{SUSY}$, 
which plays a double role: on one hand, it keeps constant values of the 
LFV Higgs decay rates (due to the non-decoupling behavior of these decays 
with $m_\text{SUSY}$) and, on the other hand, it brings down 
$\tau \to \mu \gamma$ below its experimental upper bound 
(because of the decoupling effect of LFV radiative decays with $m_\text{SUSY}$).

As we have said above, we show in Figure~\ref{BRs-tanb} the behavior of LFV branching ratios as functions of $\tan\beta$ for $\delta_{23}^{LL} =$ 0.5 (upper left panel), $\delta_{23}^{RR} =$ 0.5 (upper right panel), $\tilde \delta_{23}^{LR} =$ 0.5 (lower left panel) and $\tilde \delta_{23}^{RL} =$ 0.5 (lower right panel) with $m_A =$ 1000 GeV, $m_\text{SUSY} =$ 5 TeV and $M_2 = m_\text{SUSY}/5$. All the LFV rates have a very similar behavior with $\tan\beta$ for both $LL$ and $RR$ mixing cases and grow as $\sim (\tan\beta)^2$ for large values of $\tan\beta \gtrsim$ 10, as indicated in the previous paragraphs. In contrast, the LFV rates present a decreasing behavior with $\tan\beta$ in the $LR$ and $RL$ cases, which are identical. BR$(\tau \to \mu \gamma)$ and BR$(h \to \tau \mu)$ go approximately as $\sim (\tan\beta)^{-2}$ while BR$(H \to \tau \mu)$ BR$(A \to \tau \mu)$ grow around two orders of magnitude from $\tan\beta =$ 1 to $\tan\beta =$ 5, and from this value decrease in the same way as $\tau \to \mu \gamma$ and $h \to \tau \mu$. Therefore, within the large $\tan\beta$ regime ($\tan\beta \gtrsim$ 10), in the $LL$ and $RR$ mixing cases the LFV rates grow as $\sim (\tan\beta)^2$ whilst in the $LR$ and $LR$ ones these rates present the opposite behavior and decrease as $\sim (\tan\beta)^{-2}$.

\begin{figure}[t!]
\begin{center}
\begin{tabular}{cc}
\includegraphics[width=80mm]{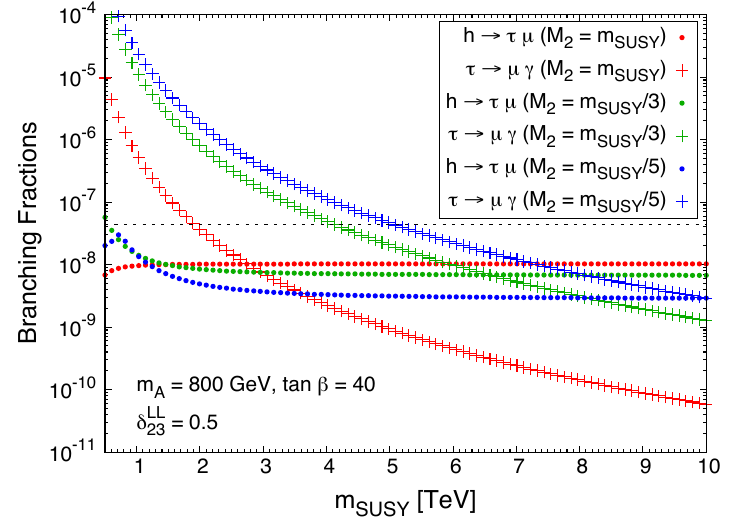} &
\includegraphics[width=80mm]{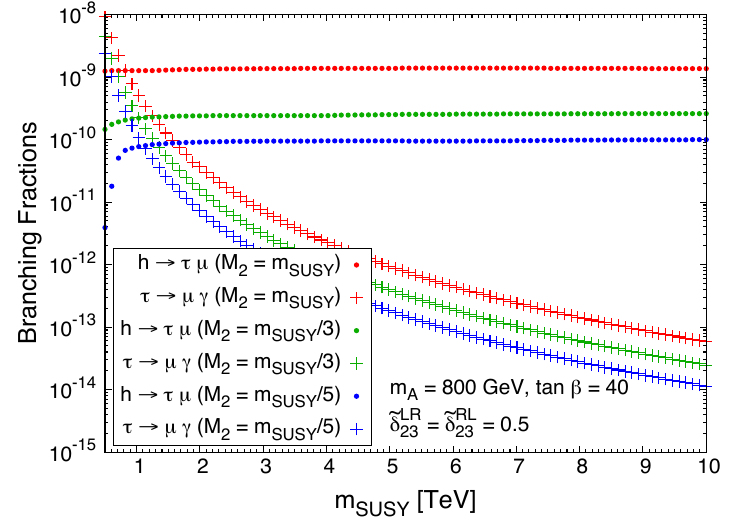} \\
\includegraphics[width=80mm]{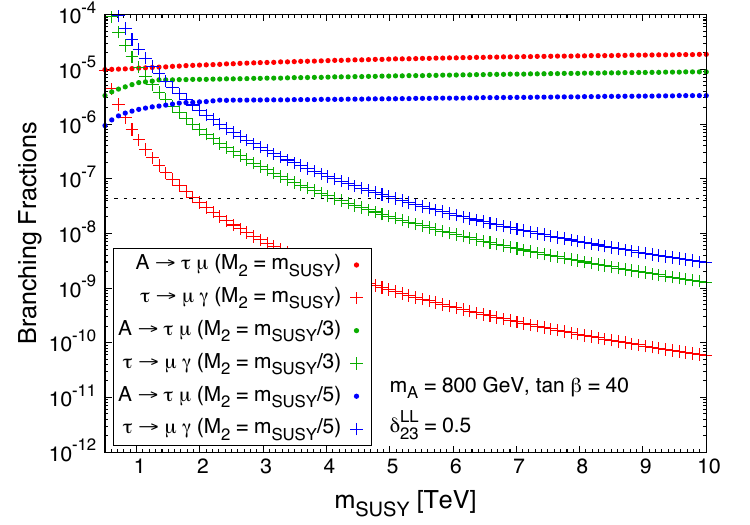} &
\includegraphics[width=80mm]{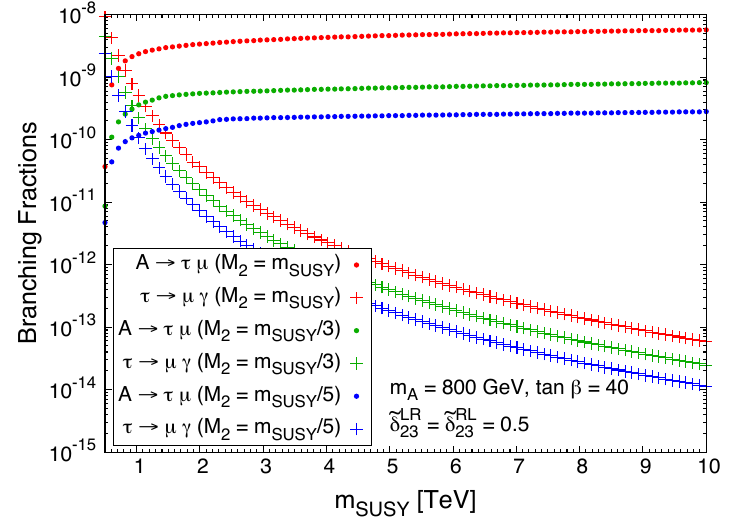} 
\end{tabular}
\caption{Sensitivity to $M_2$: LFV Higgs decay rates (dots) and BR($\tau \to \mu \gamma$) (crosses) 
as functions of $m_\text{SUSY}$ with $\delta_{23}^{LL} = 0.5$ (left panels) 
and $\tilde \delta_{23}^{LR} = \tilde \delta_{23}^{RL} = 0.5$ (right panels) for different choices of 
$M_2$: $M_2 = m_\text{SUSY}$ (in red), $M_2 = \frac{1}{3} m_\text{SUSY}$ 
(in green) and $M_2 = \frac{1}{5} m_\text{SUSY}$ (in blue). The
results for $H$ (not shown) are nearly identical to those of $A$.
In each case, the other flavor changing deltas are set to zero. 
In all panels, $m_A =$ 800 GeV, $\tan\beta =$ 40 and the other MSSM 
parameters are set to the values reported in the text. The horizontal dashed 
line denotes the current experimental upper bound for $\tau \to \mu \gamma$ 
channel, BR($\tau \to \mu \gamma$) $< 4.4 \times 10^{-8}$~\cite{Aubert:2009ag}.}\label{BRs-M2}
\end{center}
\end{figure}

Now, we are interested in investigating if other choices of $M_2$ alter these 
previous results. For this purpose, we have plotted in Figure~\ref{BRs-M2} the 
predictions of BR($h \to \tau \mu$) (dots in upper panels), BR($A \to \tau \mu$) 
(dots in lower panels) and BR($\tau \to \mu \gamma$) (crosses in all panels) 
as functions of $m_\text{SUSY}$ for different values of $a$ 
(see Eq.~(\ref{M2mu})), $a = 1$ (in red), $a = \frac{1}{3}$ (in green) 
and $a = \frac{1}{5}$ (in blue), with $\delta_{23}^{LL} = 0.5$ (left panels) 
and $\tilde \delta_{23}^{LR} = \tilde \delta_{23}^{RL} = 0.5$ (right panels). The
results for the $H \to \tau \mu$ channel are nearly identical to those of $A \to \tau \mu$ and not shown here for shortness. In the case of $LL$ mixing,
 all the LFV Higgs rates, which present the same behavior with $a$, increase 
 around a factor of 7 from $a = \frac{1}{5}$ to $a = 1$, while 
the $\tau \to \mu \gamma$ rates present the opposite behavior with $a$, decreasing in a factor about 40 for the same values of $a$. Therefore, if $\delta_{23}^{LL}$ is the responsible for the slepton mixing, and for the explored interval $1/5 \le a \le 1$, the larger $M_2$ is (and consequently $M_1$ and $\mu$), the larger the LFV Higgs branching ratios are and the lower BR($\tau \to \mu \gamma$) is. 
In the 
$LR$-mixing case we see that again BR($h, H, A \to \tau \mu$) rise as $a$ 
grows and the enhancement is by a large factor of about 15 by 
changing $a = \frac{1}{5}$ to $a = 1$. In contrast to the $LL$ case, 
BR($\tau \to \mu \gamma$) also increases with $a$ for $LR$ mixing, 
although softer than the LFV Higgs rates. In summary, we learn from Figure~\ref{BRs-M2} 
that the best choice, for a fixed delta parameter, in order 
to obtain the largest LFV Higgs rates is $M_2 = m_\text{SUSY}$. However, 
we must be very careful, because it is possible that these large rates 
are excluded by the $\tau \to \mu \gamma$ upper bound, depending basically 
on the specific values of $\delta_{23}^{LL}$, $\tilde \delta_{23}^{LR}$ and 
$\tan \beta$.

\begin{figure}[t!]
\begin{center}
\begin{tabular}{cc}
\includegraphics[width=80mm]{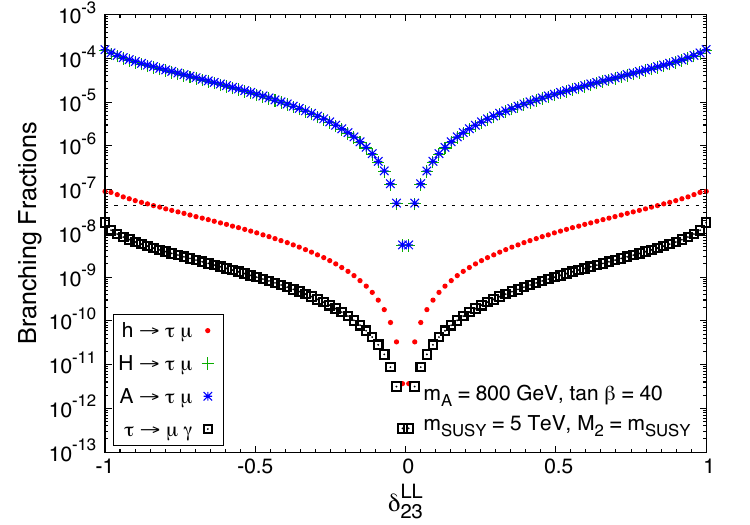} &
\includegraphics[width=80mm]{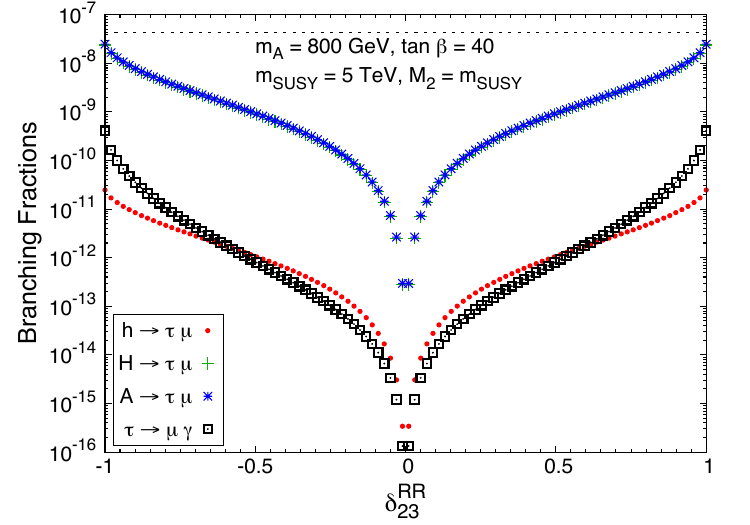} \\
\includegraphics[width=80mm]{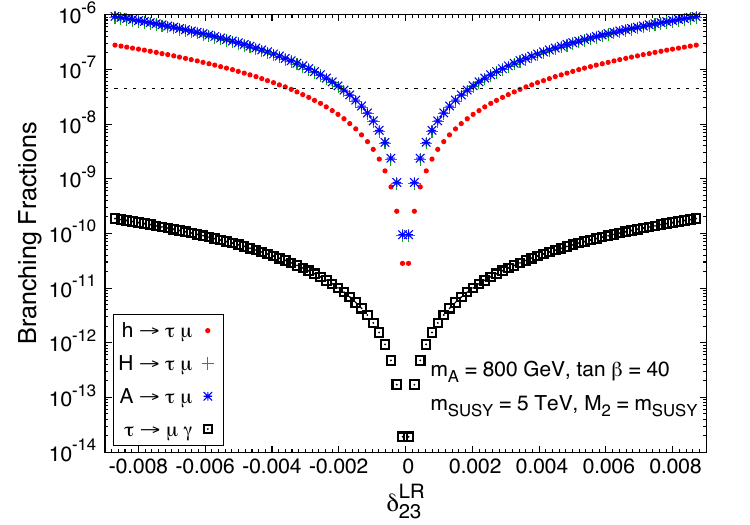} &
\includegraphics[width=80mm]{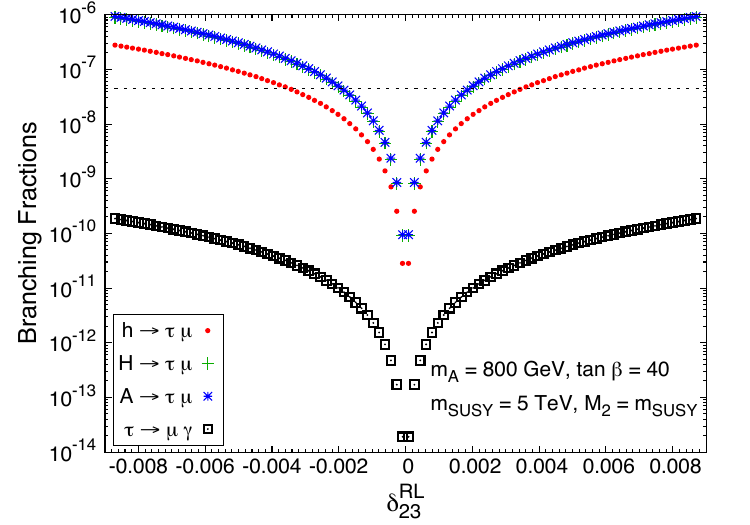}

\end{tabular}
\caption{BR($h \to \tau \mu$), BR($H \to \tau \mu$), BR($A \to \tau \mu$) and 
BR($\tau \to \mu \gamma$) as functions of $\delta_{23}^{LL}$ 
(upper left panel), $\delta_{23}^{RR}$ (upper right panel), $\delta_{23}^{LR}$ 
(lower left panel) and $\delta_{23}^{RL}$ (lower right panel). In each case, 
the other flavor changing deltas are set to zero. In all panels, $m_A =$ 800 GeV, $\tan\beta = 40$, $m_\text{SUSY} =$ 5 TeV and the other MSSM parameters are set to the values reported in the text, with $M_2 = m_\text{SUSY}$. The horizontal dashed line denotes the current experimental upper bound for $\tau \to \mu \gamma$ channel, BR($\tau \to \mu \gamma$) $< 4.4 \times 10^{-8}$~\cite{Aubert:2009ag}.}\label{BRs-deltas}
\end{center}
\end{figure}

In order to look into the largest values of $\delta_{23}^{AB}$ allowed by data 
for the choice $M_2 = m_\text{SUSY}$, we show in Figure~\ref{BRs-deltas} the 
results of the branching fractions for the LFV Higgs decays into 
$\tau \mu$ and the related LFV radiative decay $\tau \to \mu \gamma$ as 
functions of the four deltas considered along this work. We have fixed in these
plots $\tan\beta=40$. For completeness, 
we have also presented here the results for $\delta_{23}^{RR}$, which are 
irrelevant for the present work since all the branching ratios obtained are 
extremely small to be detectable at the LHC, and the $\delta_{23}^{RL}$ 
results, which are identical to the $\delta_{23}^{LR}$ ones. The plots in 
Figure~\ref{BRs-deltas} show the expected growing of the LFV rates with the 
$|\delta_{23}^{AB}|$'s, and all of them are clearly symmetric 
$\delta_{23}^{AB} \to -\delta_{23}^{AB}$. On the upper left panel we have 
the results for the $LL$ case and it is clear that all the values of 
$\delta_{23}^{LL}$, from $-1$ to 1, are allowed by data, due to the large 
suppression that $\tau \to \mu \gamma$ suffers for $m_\text{SUSY} =$ 5 TeV. 
The predictions for $H \to \tau \mu$ (green crosses) are indistinguishable 
from $A \to \tau \mu$ ones (blue asterisks), which are superimposed in these 
plots. One can reach values of BR($h \to \tau \mu$) $\simeq 10^{-7}$ and 
BR($H, A \to \tau \mu$) $\simeq 2 \times 10^{-4}$ at the most for 
$\delta_{23}^{LL} = \pm 1$. The predictions for the LFV rates as functions 
of $\delta_{23}^{LR}$ are presented on the lower left panel of 
Figure~\ref{BRs-deltas}. In this case, all the values of 
$\left|\delta_{23}^{LR}\right|$ are allowed by the $\tau \to \mu \gamma$ 
upper bound and the largest value of $\left|\delta_{23}^{LR}\right| \simeq$ 
0.009 (which corresponds to $\tilde \delta_{23}^{LR} \simeq$ 10) gives rise 
to a branching fraction of $3 \times 10^{-7}$ for the $h \to \tau \mu$ 
channel, while BR($H, A \to \tau \mu$) reach values of $1 \times 10^{-6}$. The
low rates in the $h \to \tau \mu$ channel for this $LR$-mixing case can be 
notably increased, as we have
said previously, by assuming a lower $\tan\beta$ value closer to the 
low $\tan\beta$ region with $\tan\beta \lesssim$ 5. 

\begin{figure}[t!]
\begin{center}
\begin{tabular}{cc}
\includegraphics[width=80mm]{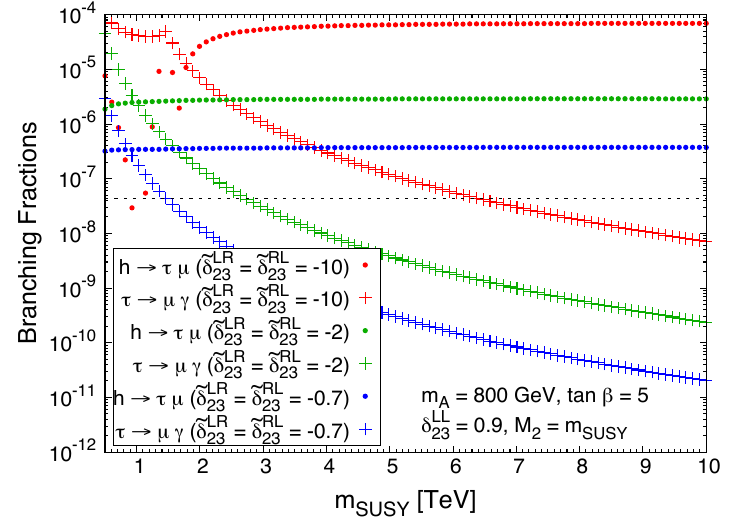} &
\includegraphics[width=80mm]{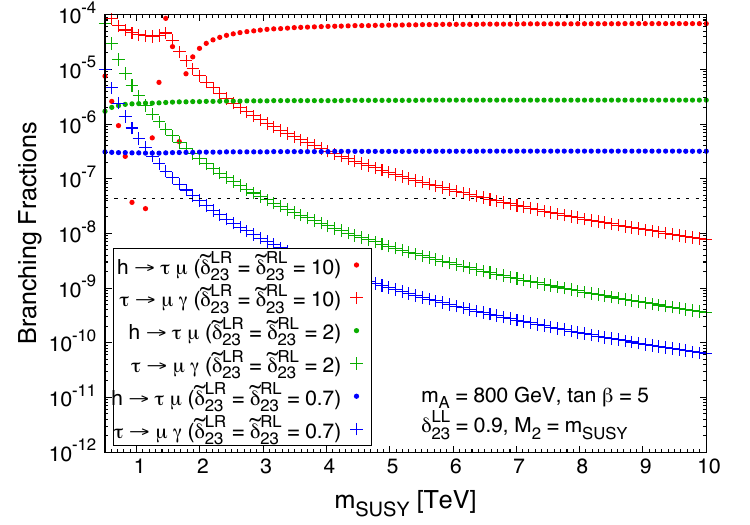} \\
\includegraphics[width=80mm]{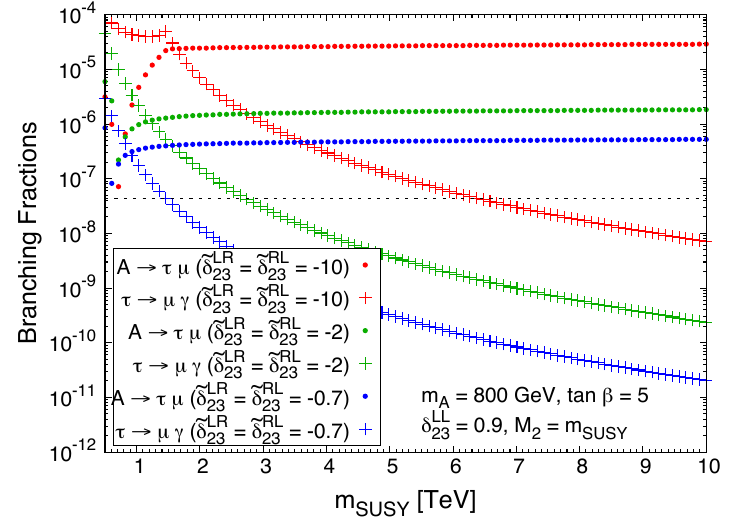} &
\includegraphics[width=80mm]{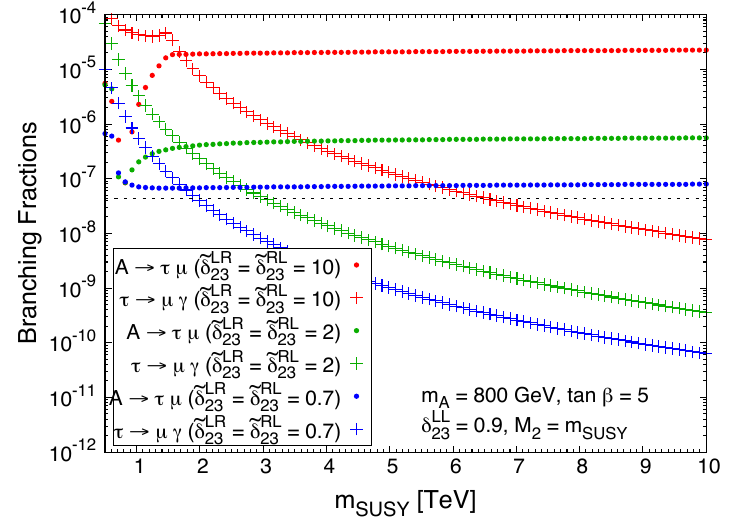} 
\end{tabular}
\caption{Sensitivity to double $LL$ and $LR$ mixing deltas: 
LFV Higgs decay rates (dots) and BR($\tau \to \mu \gamma$) (crosses) 
as functions of $m_\text{SUSY}$ with $\delta_{23}^{LL} = 0.9$ 
for different choices of
negative $LR$ mixing (left panels),
$\tilde \delta_{23}^{LR} = \tilde \delta_{23}^{RL}$: $-0.7$ (in blue), $-2$ 
(in green) and $-10$ (in red), and of positive $LR$ mixing (right panels), 
$\tilde \delta_{23}^{LR} = \tilde \delta_{23}^{RL}$: $+0.7$ (in blue), $+2$ 
(in green) and $+10$ (in red). The
results for $H$ (not shown) are nearly identical to those of $A$. 
In each case, the other flavor changing deltas are set to zero. 
In both panels, $m_A =$ 800 GeV, $\tan\beta =$ 5, $M_2 = m_\text{SUSY}$ and the other 
MSSM parameters are set to the values reported in the text. 
The horizontal dashed line denotes the current experimental 
upper bound for $\tau \to \mu \gamma$ channel, 
BR($\tau \to \mu \gamma$) $< 4.4 \times 10^{-8}$~\cite{Aubert:2009ag}.}
\label{BRs-deltatildeLRRL}
\end{center}
\end{figure}

Finally, we have studied the possibility of switching on several deltas 
simultaneously. Specifically, we have fixed $\delta_{23}^{LL} = 0.9$ and 
considered different choices of 
$\tilde \delta_{23}^{LR} = \tilde \delta_{23}^{RL}$
with either negative values:  
($-0.7$, $-2$ and $-10$), or positive values ($+0.7$,
$+2$ and $+10$). 
The results are depicted in Figure~\ref{BRs-deltatildeLRRL} for the case of low 
$\tan\beta =$ 5 that is the most interesting one since the $LL$ and $LR$
(and $RL$) contributions are of similar size and their interferences can be
relevant for some regions of the parameter space.    
As expected, the four LFV decay rates increase as 
$|\tilde \delta_{23}^{LR}| = |\tilde \delta_{23}^{RL}|$ grows, 
and they are slightly
higher than for single $LL$ or $LR$ mixings. The most 
important conclusion in this case is that we are able to obtain large 
branching ratios for all the three LFV Higgs decays, reaching values close 
to $10^{-4}$ for $h$ and about $3\times 10^{-5}$ for $A$ and $H$, if 
$\tilde \delta_{23}^{LR} = \tilde \delta_{23}^{RL} = \pm 10$. By comparing the
results for negative versus positive $LR$ mixings, we also learn from this
figure that there are not relevant differences. The LFV Higgs decay rates for 
negative mixings
are slightly higher than the corresponding rates for positive mixings, and this
difference is more visible in the $A$ and $H$ LFV decays than in the $h$ LFV 
decay. It should also be noted that the rates for $\tau \to \mu \gamma$ decays
go the other way around, namely, they are slightly larger for positive $LR$
mixings than for negative $LR$ mixings, indicating that the interference
between the $LL$ and $LR$ contributions must be of opposite sign in the LFV 
Higgs decays versus the $\tau \to \mu \gamma$ decays.

To close this section, we can conclude from Figures~\ref{BRs-mSUSY}, 
\ref{BRs-tanb}, \ref{BRs-M2}, \ref{BRs-deltas} and~\ref{BRs-deltatildeLRRL} 
that, for the explored intervals of the parameter space, the largest LFV 
Higgs rates that are allowed by the $\tau \to \mu \gamma$ upper bound are 
obtained for the following values of the model parameters: large 
$m_\text{SUSY} \gtrsim$ 5 TeV, $M_2$ close to $m_\text{SUSY}$ and
$|\delta_{23}^{LL}|$ and $|\tilde\delta_{23}^{LR}|$ (and/or
$|\tilde\delta_{23}^{RL}|$) close to their 
maximum explored values of 1 and 10, respectively. According to these previous findings, 
in the forthcoming computation of cross sections and
event rates at the LHC, whenever we have to fix them, we will set the following 
particular reference 
model parameters: 
$m_\text{SUSY} =$ 5 TeV, $M_2 = m_\text{SUSY}$, $\delta_{23}^{LL} = 0.9$ and $\tilde \delta_{23}^{LR} = \tilde \delta_{23}^{RL} = \pm 5$ , which are approximately the largest allowed values by the metastability bounds. 
The corresponding predictions for other choices of $\delta_{23}^{LL}$, 
$\tilde \delta_{23}^{LR}$, $\tilde \delta_{23}^{RL}$, $M_2$, $m_\text{SUSY}$ 
and $\tan\beta$ can be easily derived from these first five figures.

\section{Results for the LFV event rates at the LHC}
\label{LFVxsections}

In this section we present the results of the LFV event 
rates at the LHC which are mediated by the production of neutral MSSM Higgs 
bosons and their subsequent LFV decays into $\tau \mu$. The production cross 
sections of the neutral Higgs bosons are calculated here by means of the 
code {\tt FeynHiggs}~\cite{FeynHiggs}. For low values of $\tan\beta$, 
the production cross sections of the three neutral MSSM Higgs bosons are 
dominated by gluon fusion. By contrast, for moderate and large values of 
$\tan\beta$ ($\gtrsim 10$), the production cross sections of $H$ and $A$ Higgs 
bosons 
via bottom-antibottom quark annihilation become the dominant ones, 
while the $h$ production cross section is still dominated by gluon fusion. 
In the following, we consider center-of-mass energies at the LHC 
of $\sqrt{s} =$ 8 TeV, with a total integrated luminosity of ${\cal L} =$ 25 fb$^{-1}$, and $\sqrt{s} =$ 14 TeV, with ${\cal L} =$ 100 fb$^{-1}$, and focus on the two cases 
with the largest LFV Higgs decay rates, with either $LL$ or $LR$ or both 
slepton 
$\tau-\mu$ mixings. Although we do not expect any competitive background 
to these singular LFV signals at the LHC, a more realistic and devoted 
study of the potential backgrounds should be done, but this is beyond the 
scope of this paper.

\begin{figure}[t!]
\begin{center}
\begin{tabular}{cc}
\includegraphics[width=80mm]{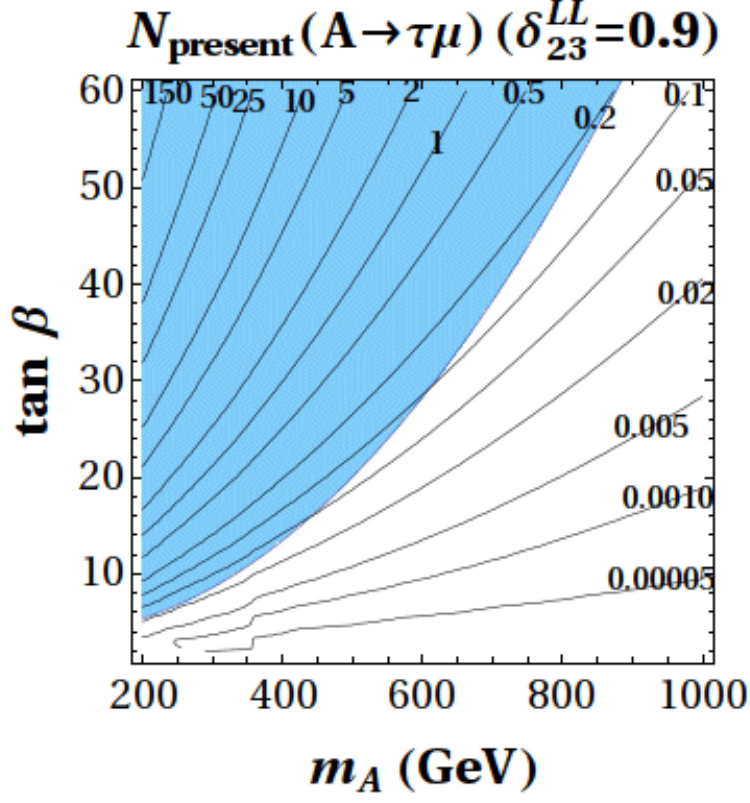} &
\includegraphics[width=80mm]{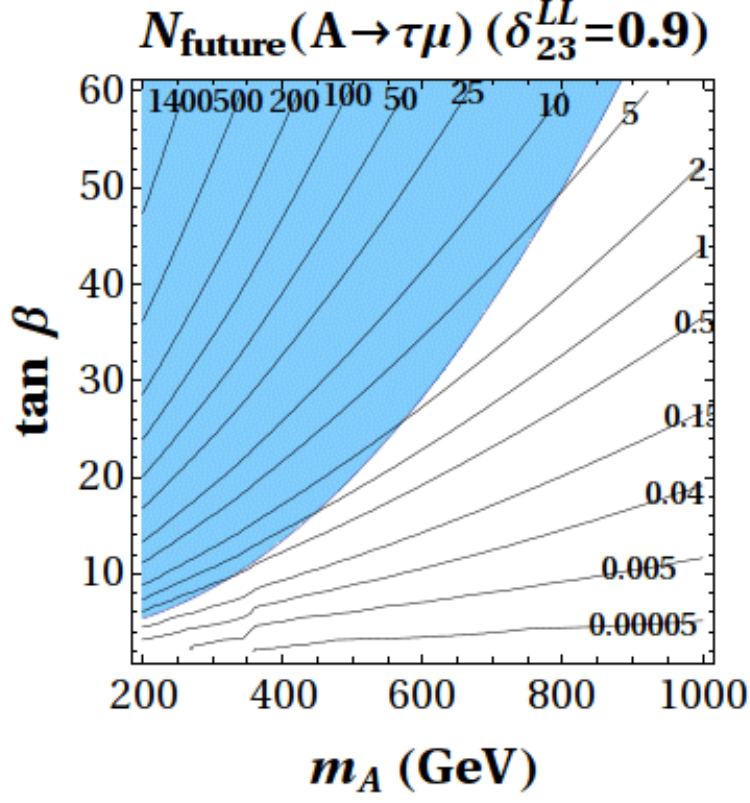}
\end{tabular}
\caption{Number of expected LFV events in the ($m_A,\tan\beta$) plane 
from $A \to \tau \mu$ for $\delta_{23}^{LL} =$ 0.9 and 
$m_\text{SUSY} =$ 5 TeV. Left panel: present phase of the 
LHC with $\sqrt{s} =$ 8 TeV and ${\cal L} =$ 25 fb$^{-1}$. 
Right panel: future phase of the LHC with $\sqrt{s} =$ 14 TeV and 
${\cal L} =$ 100 fb$^{-1}$. In both panels the other MSSM parameters 
are set to the values reported in the text, with $M_2 = m_\text{SUSY}$. 
The shaded blue areas are excluded by CMS 
searches~\cite{CMS-PAS-HIG-12-050}. The results for $H$ (not shown)
are nearly equal to these ones for $A$.}\label{NeventsA-mAtanb}
\end{center}
\end{figure}

Once we have set up the most relevant parameters for the present 
study of LFV at the LHC, which are the two flavor mixing 
deltas $\delta_{23}^{LL}$ and $\delta_{23}^{LR}$ 
(and correspondingly $\delta_{23}^{RL}$), $m_\text{SUSY}$ 
and $\tan\beta$, we will present next the results for the final rates 
at the LHC, both in the present and future phases, in the most convenient 
way for comparison with future experimental analysis, namely, in the 
($m_A,\tan\beta$) plane. Since the results
for the $H \to \tau \mu$ event rates turn out to be nearly equal 
to those of the 
$A \to \tau \mu$ ones, we will not exhibit them here for shortness. Thus we will focus
on the LFV rates of $h$ and $A$ decays. In the following plots we have
also specified the areas of the $(m_A,\tan\beta)$ plane (blue areas) that  
are excluded by 
the recent CMS searches for MSSM neutral Higgs bosons decaying to 
$\tau {\bar \tau}$ pairs in the so-called $m_h^\text{max}$ 
scenario~\cite{CMS-PAS-HIG-12-050}. All the predictions shown next are allowed
by the present $\tau \to \mu \gamma$ upper bound. 

We start this analysis with the $LL$ case and plot in 
Figure~\ref{NeventsA-mAtanb} the number of events expected in the 
($m_A,\tan\beta$) plane for the $A \to \tau \mu$ channel with 
$\delta_{23}^{LL} =$ 0.9 and $m_\text{SUSY} =$ 5 TeV, considering both 
the present and future LHC phases (left and right panels, respectively). 
 On the other hand, the $h \to \tau \mu$ channel (not shown) cannot supply any 
 significant signal 
 at the LHC in this $LL$ case, due to its very small branching ratios, unless extremely large total 
 integrated luminosities were considered (larger than 500 fb$^{-1}$). 
 Due to the CMS exclusion region in the ($m_A,\tan\beta$) 
 plane~\cite{CMS-PAS-HIG-12-050},  
 it is evident that we cannot expect any LFV event from neither 
 the $h \to \tau \mu$
 channel nor the $A,H\to \tau \mu$ channels in the present phase 
 of the LHC if the unique responsible for $\tau-\mu$ mixing is the 
 $\delta_{23}^{LL}$ parameter and $|\delta_{23}^{LL}| <$ 1. 
 The event rates from $A, H \to \tau \mu$ for the future phase of the LHC 
 are more promising, as shown 
 on the right panel of Figure~\ref{NeventsA-mAtanb}. For instance, 
 for $m_A \simeq$ 450 GeV and $\tan\beta \simeq$ 15, we could expect
 at least 1 event, and up to 5 for larger values of $m_A$ and 
 values of $\tan\beta \gtrsim$ 50.
 
\begin{figure}[t!]
\begin{center}
\begin{tabular}{cc}
\includegraphics[width=80mm]{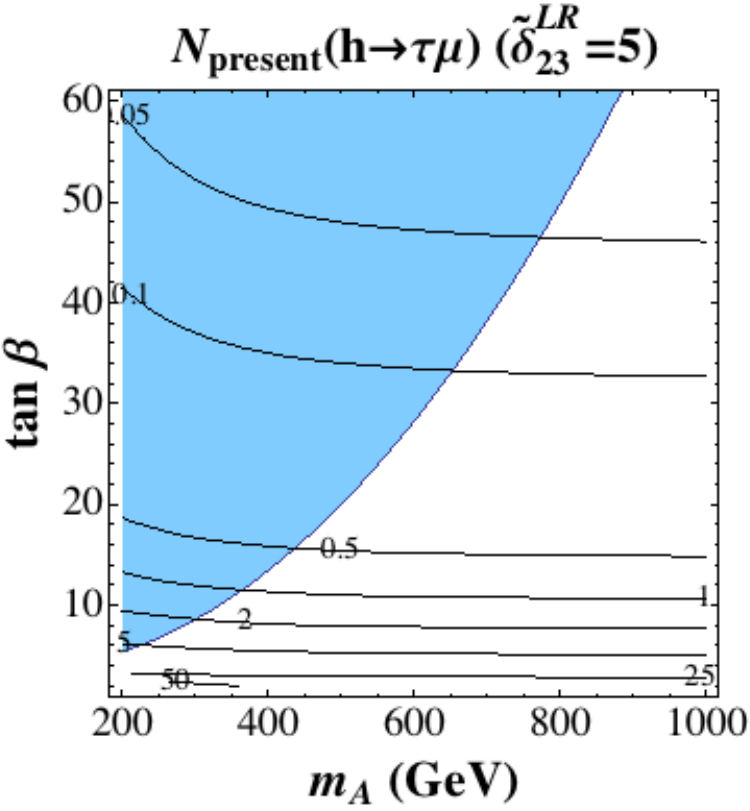} &
\includegraphics[width=80mm]{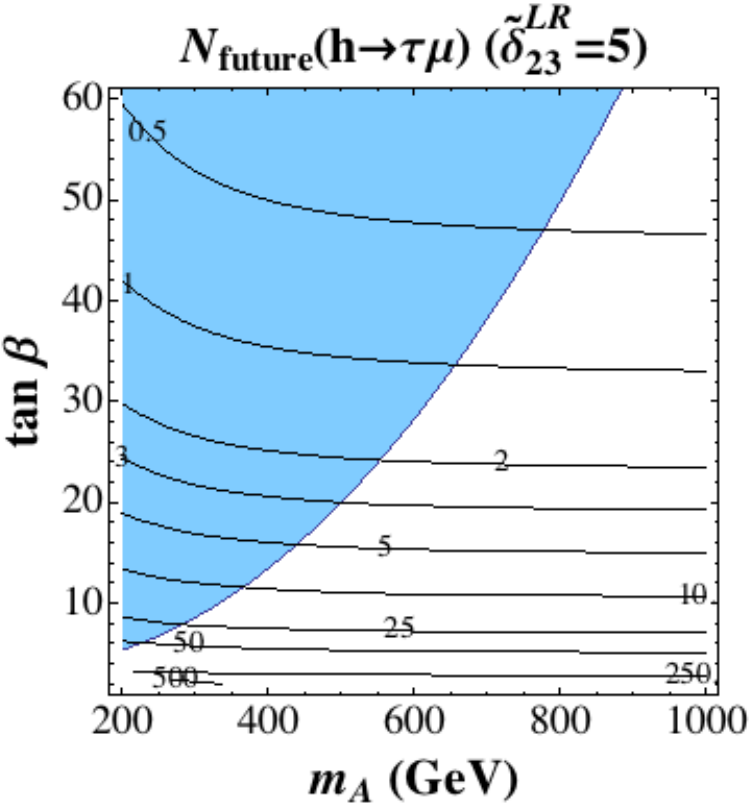}
\end{tabular}
\caption{Number of expected LFV events in the ($m_A,\tan\beta$) plane from 
$h \to \tau \mu$ for $\tilde \delta_{23}^{LR} = \tilde \delta_{23}^{RL} =$ 5 
and $m_\text{SUSY} =$ 5 TeV. Left panel: present phase of the LHC 
with $\sqrt{s} =$ 8 TeV and ${\cal L} =$ 25 fb$^{-1}$. Right panel: future phase of the LHC with $\sqrt{s} =$ 14 TeV and ${\cal L} =$ 100 fb$^{-1}$. In both panels the other MSSM parameters are set to the values reported in the text, with $M_2 = m_\text{SUSY}$. The shaded blue areas are excluded by CMS searches~\cite{CMS-PAS-HIG-12-050}.}\label{Neventsh-mAtanb-deltaLR}
\end{center}
\end{figure}

\begin{figure}[t!]
\begin{center}
\begin{tabular}{cc}
\includegraphics[width=80mm]{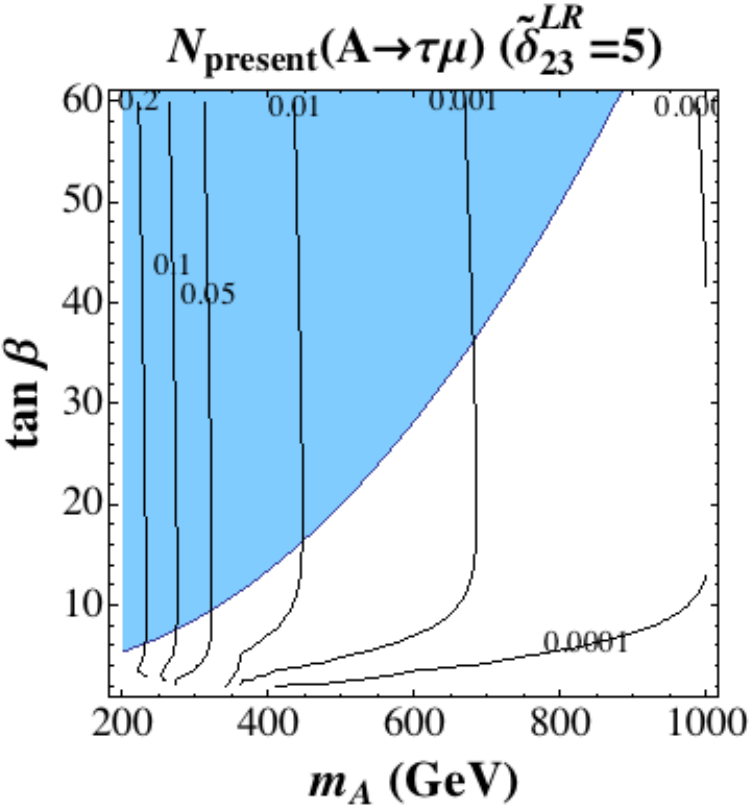} &
\includegraphics[width=80mm]{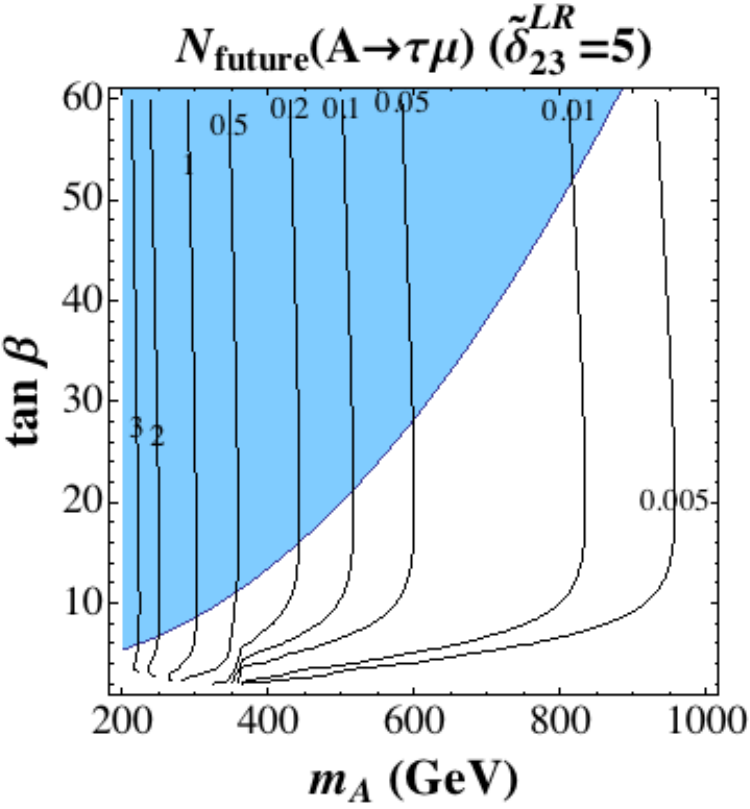}
\end{tabular}
\caption{Number of expected LFV events in the ($m_A,\tan\beta$) 
plane from $A \to \tau \mu$ for 
$\tilde \delta_{23}^{LR} = \tilde \delta_{23}^{RL} =$ 5 and 
$m_\text{SUSY} =$ 5 TeV. 
Left panel: present phase of the LHC with $\sqrt{s} =$ 8 TeV 
and ${\cal L} =$ 25 fb$^{-1}$. Right panel: future phase of 
the LHC with $\sqrt{s} =$ 14 TeV and ${\cal L} =$ 100 fb$^{-1}$. 
In both panels the other MSSM parameters are set to the values reported 
in the text, with $M_2 = m_\text{SUSY}$. The shaded blue areas 
are excluded by CMS searches~\cite{CMS-PAS-HIG-12-050}. 
The results for $H$ (not shown)
are nearly equal to these ones for $A$.}\label{NeventsA-mAtanb-deltaLR}
\end{center}
\end{figure}

 Next we analyze the results for the case of $LR$ and $RL$ mixings 
in the 
($m_A,\tan\beta$) plane. Figures~\ref{Neventsh-mAtanb-deltaLR} 
and~\ref{NeventsA-mAtanb-deltaLR} summarize the results for the 
$h \to \tau \mu$ and $A \to \tau \mu$ channels, respectively, 
in the present and future LHC stages. 

On the left panel of Figure~\ref{Neventsh-mAtanb-deltaLR}, 
where the number of expected events from the $h \to \tau \mu$ channel 
in the present phase of the LHC are shown as a function of 
$m_A$ and $\tan\beta$, for $m_\text{SUSY} =$ 5 TeV and 
$\tilde \delta_{23}^{LR} = \tilde \delta_{23}^{RL} =$ 5, we see again
that the maximum allowed number of events are obtained in the low 
$\tan\beta$ region. Tens of events are expected, up to 50 for 
$\tan\beta \lesssim$ 3, in all the studied $m_A$ interval. 
In any case, in all the allowed region the number of predicted 
events are softly dependent on $m_A$ and at least 
one event is obtained, even for large values of $m_A$ and 
$\tan\beta \lesssim$ 10. On the right panel 
of Figure~\ref{Neventsh-mAtanb-deltaLR}, the predictions for the 
$h \to \tau \mu$ channel, in the future LHC phase with a 
center-of-mass energy of $\sqrt{s} =$ 14 TeV and a total integrated 
luminosity of ${\cal L} =$ 100 fb$^{-1}$, show the same behavior 
with respect to the two pair of parameters as on the left panel but 
with an increase in the number of events of around one 
order of magnitude. Again the maximum amount 
of events are for the lowest $\tan\beta$ values, being these 
nearly independent on $m_A$, and the rates decrease as we raise 
$\tan\beta$, showing a small variation with respect to 
$m_A$ for the allowed region by data (in white), 
as in the previous mentioned plot. Specifically, we 
obtain up to 500 events for $\tan\beta \simeq$ 2, and between 250 and 1 events 
for the region between $\tan\beta =$ 2 and $\tan\beta =$ 35.

The corresponding results for the $A \to \tau \mu$ channel, displayed 
in Figure~\ref{NeventsA-mAtanb-deltaLR}, show a very different behavior with 
$m_A$ and $\tan\beta$ than the previous $h$ case. The number of expected 
LFV events at the LHC via $A \to \tau \mu$ decays 
diminish as $m_A$ increases, due to the suppression in the production 
cross section of a heavy pseudoscalar Higgs boson, and stay constant 
with $\tan\beta$, due mainly to the compensation between the 
growing of the $A$ production cross section via 
bottom-antibottom quark annihilation and the reduction 
of BR$(A \to \tau \mu)$ with this parameter, as previously illustrated 
in Figure~\ref{BRs-tanb}. In the present phase of the LHC we cannot 
expect any event, as shown on the left panel of 
Figure~\ref{NeventsA-mAtanb-deltaLR}. The right panel of 
Figure~\ref{NeventsA-mAtanb-deltaLR}, containing the predictions 
for the $A \to \tau \mu$ channel in the future LHC phase, shows an 
analogous behavior to that of the left panel. 
The number of expected events increase around one order of magnitude 
respect the present LHC phase, and for values of $m_A$ below 300 GeV, 
one could expect between 1 and 3 events independently on the value 
of $\tan\beta$.

\begin{figure}[t!]
\begin{center}
\begin{tabular}{cc}
\includegraphics[width=80mm]{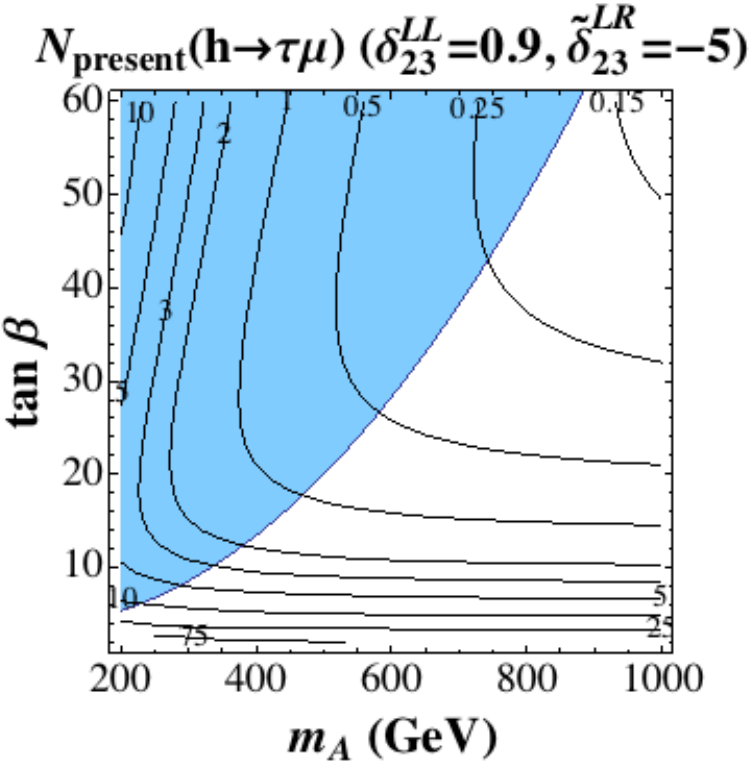} &
\includegraphics[width=80mm]{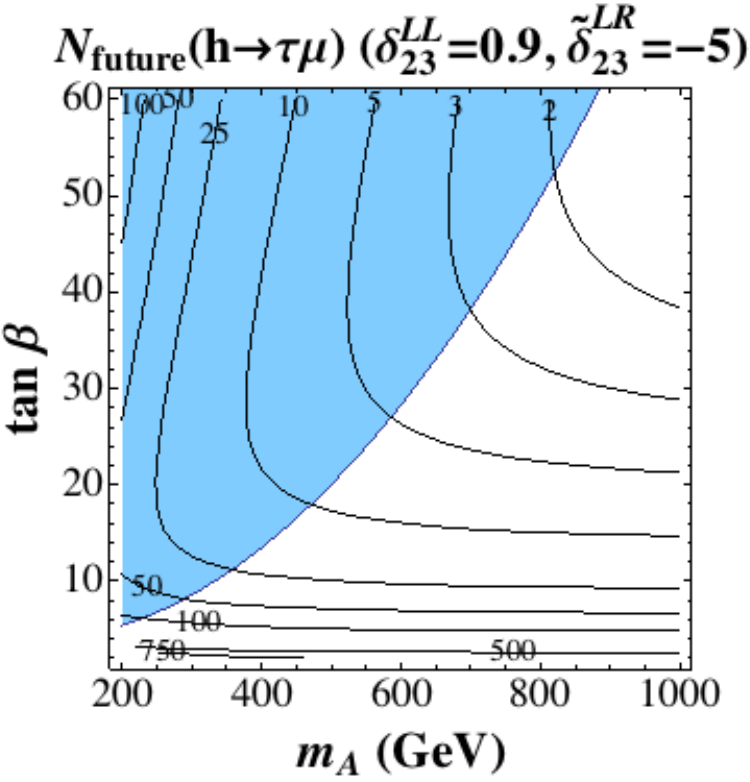}\\
\includegraphics[width=80mm]{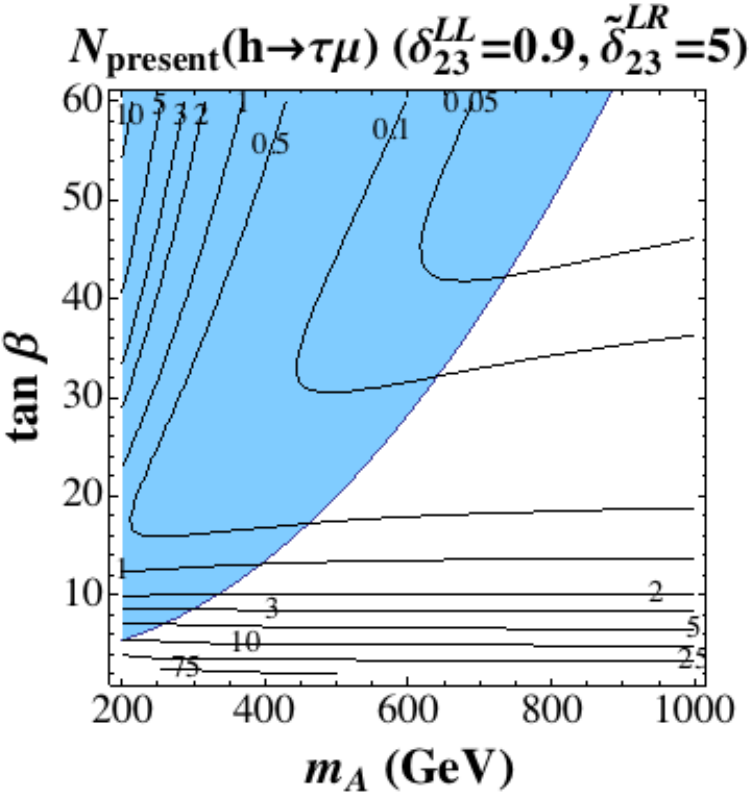} &
\includegraphics[width=80mm]{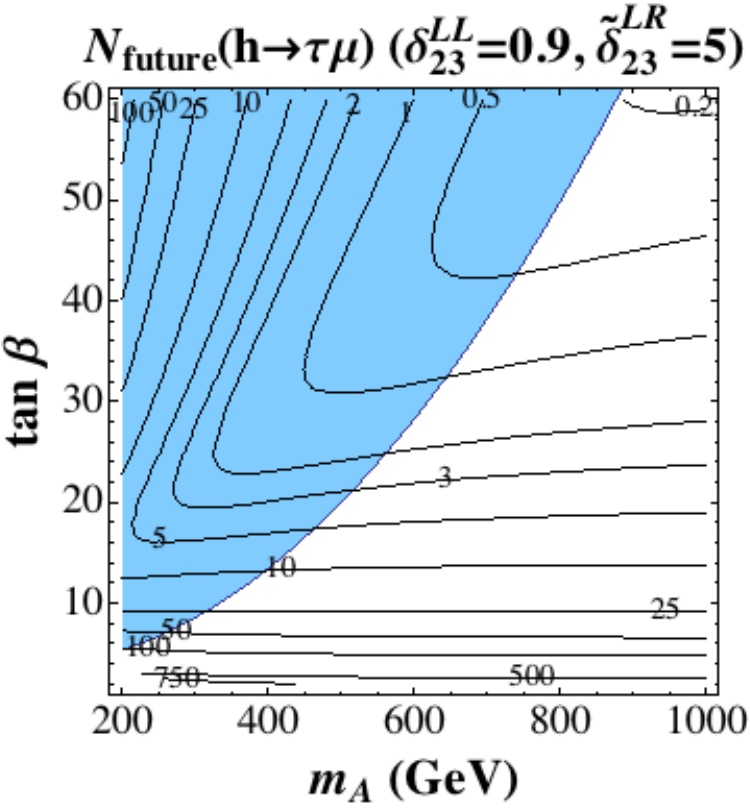}

\end{tabular}
\caption{Number of expected LFV events in the ($m_A,\tan\beta$) plane from 
$h \to \tau \mu$ for $m_\text{SUSY} =$ 5 TeV, $\delta_{23}^{LL} =$ 0.9 and 
$\tilde \delta_{23}^{LR} = \tilde \delta_{23}^{RL} =$ -5 (upper pannels) or 
$\tilde \delta_{23}^{LR} = \tilde \delta_{23}^{RL} =$ +5 (lower pannels). 
Left panels: present phase of the LHC with 
$\sqrt{s} =$ 8 TeV and ${\cal L} =$ 25 fb$^{-1}$. Right panels: future phase 
of the LHC with $\sqrt{s} =$ 14 TeV and ${\cal L} =$ 100 fb$^{-1}$. 
In all panels the other MSSM parameters are set to the values reported 
in the text, with $M_2 = m_\text{SUSY}$. The shaded blue areas are 
excluded by CMS searches~\cite{CMS-PAS-HIG-12-050}.}
\label{Neventsh-mAtanb-deltaLLdeltaLR}
\end{center}
\end{figure}

\begin{figure}[t!]
\begin{center}
\begin{tabular}{cc}
\includegraphics[width=80mm]{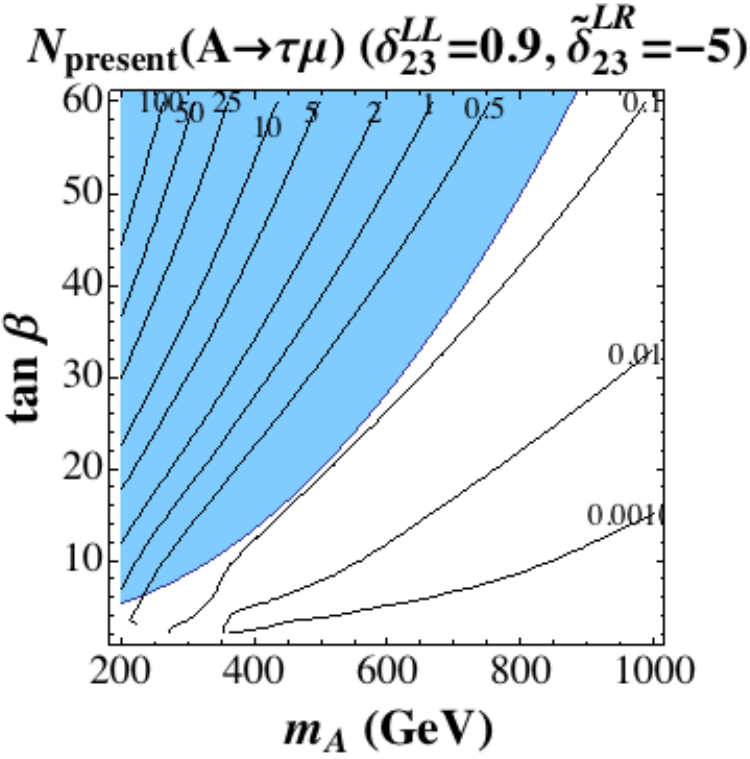} &
\includegraphics[width=80mm]{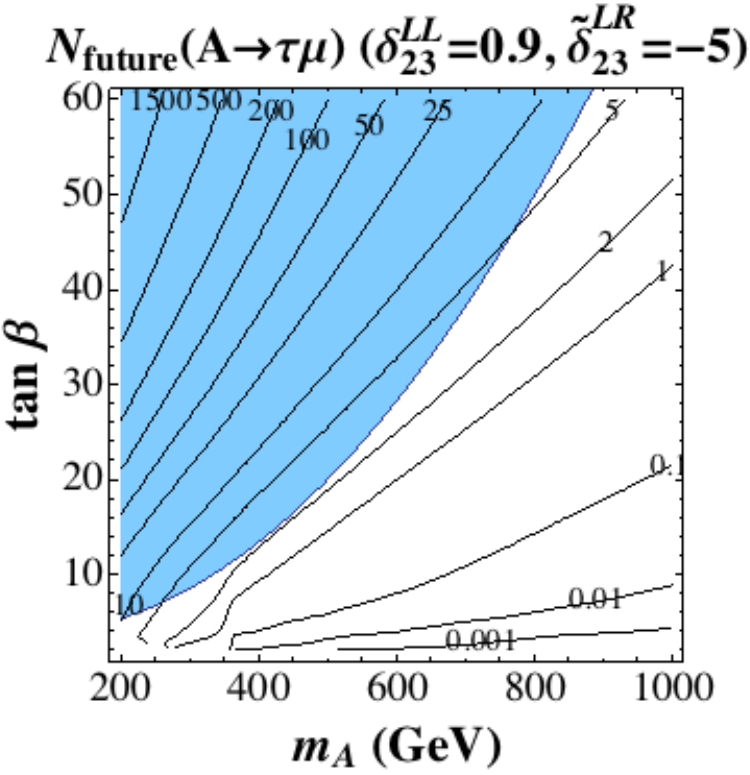}
\end{tabular}
\caption{Number of expected LFV events in the ($m_A,\tan\beta$) plane 
from $A \to \tau \mu$ for $\delta_{23}^{LL} =$ 0.9, 
$\tilde \delta_{23}^{LR} = \tilde \delta_{23}^{RL} =$ -5 and 
$m_\text{SUSY} =$ 5 TeV. Left panel: 
present phase of the LHC with $\sqrt{s} =$ 8 TeV and ${\cal L} =$ 
25 fb$^{-1}$. Right panel: future phase of the LHC with $\sqrt{s} =$ 14 TeV 
and ${\cal L} =$ 100 fb$^{-1}$. In both panels the other MSSM parameters 
are set to the values reported in the text, with $M_2 = m_\text{SUSY}$. 
The shaded blue areas are excluded by CMS searches~\cite{CMS-PAS-HIG-12-050}.
The results for $H$ (not shown)
are nearly equal to these ones for $A$.}
\label{NeventsA-mAtanb-deltaLLdeltaLR}
\end{center}
\end{figure}

Finally, the results for the $h \to \tau \mu$ and $A \to \tau \mu$ channels 
in the double $LL$ and $LR$ mixing case are summarized in 
Figures~\ref{Neventsh-mAtanb-deltaLLdeltaLR} 
and~\ref{NeventsA-mAtanb-deltaLLdeltaLR}, respectively. 
 
The predictions for the contour lines of $h \to \tau \mu$ event rates  
(Figure~\ref{Neventsh-mAtanb-deltaLLdeltaLR}) show a clear different 
pattern than in the previous cases of single deltas. In both LHC phases, 
we achieve 
an increase in the number of events respect to the single $LL$ and 
$LR$ (and/or
$RL$) mixing cases. The most interesting numbers are at the lower part 
of these plots with $\tan \beta < 10$. We find as large as 75 LFV events 
in the present phase 
of the 
LHC for very low values of $\tan\beta\simeq $ 2 and 
$m_A \lesssim$ 600 GeV.  
In the future LHC phase we predict up 
to 750 events, 
for $\tan\beta \simeq$ 2 and $m_A \lesssim$ 450 GeV. It should be also noted that 
these conclusions apply to
both choices for the 
$LR/RL$ mixings of -5
and +5, as can be seen in Figure~\ref{Neventsh-mAtanb-deltaLLdeltaLR}. 
The only notable 
differences that we find between the results of these two cases are in the
slight different patterns of the contour lines, signalling a small different
sensitivity to $m_A$ and/or $\tan\beta$. Also one can appreciate in these plots 
that one gets a bit 
lower rates for +5 than for -5, in agreement with our previous findings 
reported in Figure~\ref{BRs-deltatildeLRRL}  

The number of events for the $A \to \tau \mu$ channel in the double $LL$ and 
$LR$ mixing case are displayed in 
Figure~\ref{NeventsA-mAtanb-deltaLLdeltaLR}. The $LL$ mixing is set here 
to 0.9 and the $LR=RL$ mixings are fixed to -5. 
As expected, the predicted event rates increase as $\tan\beta$ grows and are 
reduced as $m_A$ gets bigger, due to the suppression in the production cross 
section of a heavy pseudoscalar Higgs boson. However, 
for the present LHC phase one cannot say much about this channel, 
since all the $m_A-\tan\beta$ regions which could 
produce 
a relevant number of 
LFV 
events are excluded at present by CMS searches. We find at the most one event 
in the tiny low left corner of the allowed region in this $(m_A,\tan\beta)$ 
plot. In contrast,
the predictions in the future phase of the LHC are more promising. 
We predict up to about 5 LFV events for large values of $m_A$ and 
$\tan\beta$ and up to 10 in the low $\tan\beta$ region with 
$m_A \simeq$ 200 GeV. Similar conclusions are found for the case 
of positive $LR$=+5 mixing (not shown). The shape of the contour lines 
in this case are slightly
modified at low $\tan\beta$ but with no relevant implications in terms of event
rates. 

\section{Conclusions}
\label{conclusions}

The discovery of a new Higgs-like particle at the LHC is concentrating a 
lot of efforts in studying its properties, couplings and decays, in order to 
investigate if there is new physics behind it. In that sense, lepton flavor 
violating Higgs decays are of special interest, since they would clearly 
imply the existence of physics beyond the SM. We have discussed in this 
paper the possibility of obtaining sizeable LFV Higgs rates, induced by 
heavy SUSY, and detectable at the LHC. In particular, we have studied in 
detail the most interesting LFV decays of the three neutral MSSM Higgs 
bosons: $h \to \tau \mu$, $H \to \tau \mu$ and $A \to \tau \mu$. We have 
shown that these three channels present a non-decoupling behavior with 
$m_\text{SUSY}$. On one hand, independently if $\delta_{23}^{LL}$, 
$\delta_{23}^{RR}$, $\delta_{23}^{LR}$ or $\delta_{23}^{RL}$ are 
the responsible for the intergenerational mixing in the slepton sector, 
these LFV Higgs decay rates remain constant with $m_\text{SUSY}$ at large  $m_\text{SUSY}>$ 2 TeV. 
On the other hand, the related LFV radiative decay, $\tau \to \mu \gamma$, 
manifests a fast decoupling behavior with $m_\text{SUSY}$ and its rates 
are very suppressed for large values of the SUSY scale. In this work, 
we have taken advantage of these two remarkable different behaviors with 
$m_\text{SUSY}$ 
in order to reach sizeable LFV Higgs branching ratios which are yet 
allowed by the present $\tau \to \mu \gamma$ upper bound.

From our detailed analysis of the LFV Higgs decays, 
we have learned that the single $\tau-\mu$ mixing of $RR$ type in the 
slepton sector cannot by itself provide sufficiently large rates which 
could be measurable at the LHC. The situation ameliorates 
slightly if we consider the single $LL$ mixing case, with the largest rates
at the large $\tan\beta$ region and amounting  
up to about 5 LFV events for the $H, A \to \tau \mu$ channels at the 
future phase of the LHC. We find that the mixing parameter 
$\delta_{23}^{LR}$ 
(and $\delta_{23}^{RL}$) is the most relevant one and even when acting as 
single
mixing parameter already 
gives rise  
to sizeable and allowed by data LFV Higgs-mediated event rates
for sufficiently large values of 
$m_\text{SUSY} \geq 5$ TeV, where the LFV radiative 
$\tau \to \mu \gamma$ rates are suppressed below its present 
experimental upper bound. In this single $LR$-mixing case, 
we also find that the most promising channel is by far the 
$h \to \tau \mu$ decay for which we predict up to 50 events for low $\tan \beta$
in the present phase of the LHC. Regarding the $H, A \to \tau \mu$ channels, 
no events are expected in the present LHC phase if there is single $LR$ (and/or $RL$) 
mixing. The situation improves noticeably if one considers the 
future phase of the LHC. In this case, with a total integrated luminosity 
of 100 fb$^{-1}$, we predict hundreds of LFV events 
from the lightest Higgs boson decay into $\tau \mu$, as much as 
$5 \times 10^2$ events for very low values of $\tan\beta$. 
For the LFV heavy Higgs bosons decays the expectations are lower, 
but also increase in comparison with the present LHC phase, and 
a few events could be obtained depending on the values of 
$m_A$, $\tan\beta$ and $m_\text{SUSY}$. 
Finally, by considering double $LL$ and $LR$ mixings we obtain the most
interesting situation, since on one hand the rates are slightly increased with respect to
the single $LR$ mixing case and, on the other hand, the slight change in the 
sensitivity to 
$\tan\beta$ makes that larger values of this parameter than in the single $LR$
mixing case give rise also to sizeable event rates. For instance, one can get in
this double mixing case  
a few events at the present stage of the LHC even for moderate 
$\tan\beta \sim 15$ and large $m_A \geq $ 500 GeV. In the future LHC phase 
the reach to larger $\tan \beta$
values increases and one gets some event even at very large $\tan\beta\sim 40$
and $m_A \geq$ 800 GeV.
The largest rates found are in any case for $h \to \tau \mu$ and are clearly
localized at the
low $\tan\beta$ region where we predict 
for the present LHC phase up to about 75 events, and 
up to about 750 LFV events for the future LHC phase. 
In the future LHC phase, we get about 10 events at the most for the $H, A \to \tau \mu$ channels in the low
$\tan \beta$ region and 5 events at the most in the high 
$\tan \beta$ region. As a final comment, it is worth recalling that all the 
rates presented
in this work are doubled if one adds the decay events for the two possible final
states, $\tau^+ \mu^-$ and $\tau^- \mu^+$.

The encouraging results presented along this work strongly 
suggest that a dedicated search for the proposed LFV Higgs decays 
is extremely worthwhile and we believe that it should be further studied 
by the experiments at the LHC. If the SUSY mass scale is too heavy, 
as the present experiments are pointing out, and the SUSY particles cannot 
be directly reachable at the present or next future LHC energies, 
our proposal for LFV Higgs decays could provide an unique window 
to explore new physics and to find some hint of very heavy SUSY at the 
LHC.

\section*{Acknowledgments}

E.A. thanks Alejandro Szynkman for fruitful discussions and a careful reading of the manuscript. M.J.H thanks Andreas Crivellin for a
critical reading of a preliminary version of this paper and for some interesting comments and discussions. M.J.H also thanks Alberto Casas for his clarifying comments on the general bounds from vacuum instabilities, their domain of applicability and the way out from these bounds. The authors also thank Sven Heinemeyer for helpful information about the code {\tt FeynHiggs}. E.A. is financially supported by a MICINN postdoctoral fellowship (Spain), under grant No. FI-2010-0041, and partially by ANPCyT (Argentina) under grant No. PICT-PRH 2009-0054 and by CONICET (Argentina) PIP-2011. E.A. thanks IFLP-CONICET for hospitality and support. The work of M.H. and M.A.-C. was partially supported by CICYT (grant FPA2009-09017), the Comunidad de Madrid project HEPHACOS, S2009/ESP-1473, the European Union FP7 ITN INVISIBLES (Marie Curie Actions, PITN-GA-2011-289442) and also by the Spanish Consolider-Ingenio 2010 Programme CPAN (CSD2007-00042).

\bibliographystyle{unsrt}

\end{document}